\documentclass[ssy,preprint]{imsart}

\RequirePackage[OT1]{fontenc}
\usepackage{amsthm}
\usepackage{amsmath}
\usepackage{amssymb}
\usepackage{amsfonts}
\usepackage{mathrsfs}
\usepackage{centernot}
\usepackage{graphicx}
\RequirePackage[numbers]{natbib}
\RequirePackage[colorlinks,citecolor=blue,urlcolor=blue]{hyperref}

\usepackage{bbm}
\usepackage{pdflscape}
\usepackage{floatrow}
\usepackage{rotating}
\usepackage[toc,page]{appendix}
\usepackage{caption}
\usepackage{subcaption}

\startlocaldefs
\setattribute{journal}{name}{}

\newcommand{\dd}{ \text{d}} 	

\newcommand{\Exp}[1]{\text{exp}\left\{{#1}\right\}}
\endlocaldefs

\begin{document}

\begin{frontmatter}

\title{Over-the-counter market models with several assets}
\runtitle{OTC market models with several assets}


\begin{aug}
\author[a]{\fnms{Alain} \snm{B\'{e}langer}\corref{}\ead[label=e1]{alain.a.belanger@usherbrooke.ca}}
\and
\author[b]{\fnms{Gaston} \snm{Giroux}\ead[label=e2]{gasgiroux@hotmail.com}}
\and
\author[c]{\fnms{Miguel} \snm{Moisan-Poisson}\ead[label=e3]{miguel.moisan-poisson@usherbrooke.com}}

\runauthor{B\'{e}lange et al.}
\affiliation{Universit\'{e} de Sherbrooke}

\address[a]{Alain B\'{e}langer\\
Universit\'{e} de Sherbrooke\\
Facult\'{e} d'administration\\
2500, Universit\'{e} Blvd.\\
Sherbrooke, Canada, J1K 2R1\\
\printead{e1}}

\address[b]{Gaston Giroux\\
410, Vimy Street, Apt. 1\\
Sherbrooke, Canada, J1J 3M9\\
\printead{e2}}

\address[c]{Miguel Moisan-Poisson\\
Universit\'{e} de Sherbrooke\\
Facult\'{e} d'administration\\
2500, Universit\'{e} Blvd.\\
Sherbrooke, Canada, J1K 2R1\\
\printead{e3}}

\end{aug}

\begin{abstract}
We study two classes of over-the-counter markets specified by systems of ODE's, in the spirit of Duffie-G\^{a}rleanu-Pedersen \cite{Duffie2005}. We first compute the steady states for many of these ODE's. Then we obtain the prices at which  investors trade with each other at these steady states. Finally, we study the stability of the solutions of these ODE's.
\end{abstract}



\received{\smonth{8} \syear{2013}}

\end{frontmatter}


\tableofcontents

\section{Introduction}

This article addresses the question of equilibrium price formation and stability in relatively opaque over-the-counter (OTC) markets with several traded assets. The financial crisis of 2008 brought significant concerns regarding the r\^{o}le of OTC markets, particularly
from the viewpoint of global financial stability. Darrell Duffie's recent
monograph, \textit{Dark Markets} (see \citet{Duffie2012}), documents some of the modelling efforts done to understand the effects of illiquidity associated with search
and bargaining. Duffie also notes that this area is still underdeveloped in
comparison with the vast literature available on central market mechanisms.

 Our goal is to shed some light on foundational issues in asset pricing in OTC markets with several assets. In particular, we study models of OTC markets described by ODE's which happen to have a financial market (time invariant) equilibrium (that is, a steady state). In doing so, we are lead to ODE's which have not yet appeared
 in the differential equations literature. For the specialists in financial economics, it is well known that in OTC markets, an
investor who wishes to sell must search for a buyer, incurring opportunity and other
costs until one is found (see for instance \citet{Duffie2005}). For the case of one asset, the evolution of an investor's state can be described  by a system of four quadratic differential equations, an overview is given in Chapter 4 of \citet{Duffie2012}. There the author
develops a search-theoretic model of the cross-sectional distribution of asset
returns, under the hypothesis that the eagerness of the investors are the same whether they have the asset or not. Here we study the more general case with several assets for two classes of extended models which are still described by systems of quadratic differential equations, but without the particular hypothesis. One should notice that without changes of positions the system would stop after a finite time and the market would become inefficient.

For the first extended model, we do not track the particular asset an  investor wants to buy when she enters the market (it is called the non-segmented model/case); but the frequency at  which she enters the market depends on that asset. For the second model, we do keep track of the asset an investor intend to purchase (it is called the partially-segmented model/case). In both of our cases the quantities of each asset do not have to be the same. Here we study these two classes of markets in the spirit of \citet{Duffie2005}. When there is only one traded asset, as in DGP, the two cases collapse to the same  model. Unlike, DGP, we do not assume that the investors' eagerness are the same whether they own the asset or not. The departure from this assumption in DGP requires us to use techniques
from the theory of dynamical systems.

In such a framework, the first thing we need to show is the existence of a steady state (this steady state is designated, in the financial
literature, by the equilibrium (time-invariant) cross-sectional variation in the distribution of
ownership).  To gain insights on these systems out of equilibrium, we also show that each of our systems is asymptotically stable for any given number of assets in the case of non-segmented markets and for (one and) two assets in the case of partially segmented markets. We show the latter using the old criterion of Routh-Hurwitz (see, for instance, \citet{Dorf2011}). The criterion gets very steeply more difficult to handle as we increase the number of assets. (See also \citet{Grasselli2012} for another example of the use of this criterion in a financial
context.)

In Section 2, we describe our two classes of models. In Section 3.1, we show the existence of a steady steate and compute it explicitly for the non-segmented case for any given number of assets. In Section 3.2, we do the same for the case of partially-segmented markets with two assets. Then in Section 4, we obtain the prices on which the investors agreed and we give numerical exmples in section 5. Finally, in Section 6, we study the asymptotic stability of our systems.

\section{Two classes of models}

 \citet{Duffie2012} and \citet{Duffie2005} present their model of OTC market with one traded asset  as a system of four linear ODE's with two constraints which can be
reduced to a system of two differential equations
with two constraints. In this section, we describe two extensions of their model
involving $K\geq 1$ assets. Before describing each model in details, we would like to set up a few general definitions.

The set of available assets will be denoted $\mathcal{I} = \{1,...,K\}$. Investors can hold at most one unit of any asset $i\in\mathcal{I}$ and cannot short-sell. Time is treated continuously and runs forever. The market is populated by a continuum of investors. At each time, an investor is characterized by whether he owns the $i$-th asset or not, and by an intrinsic type which is either a 'high' or a 'low' liquidity state. Our interpretation of liquidity state is the same as in \citet{Duffie2005}. For example, a low-type investor who owns an asset may have a need for cash and thus wants to liquidate his position. A high-type investor who does not own an asset may want to buy the asset if he has enough cash. Through time, investors' ownerships will switch randomly because of meetings leading to trades, at a rate $\lambda_i$, and the investor's intrinsic type will change independently  via an autonomous movement. This dynamics of investor's type change is modeled by a (non-homogeneous) continuous-time Markov chain $Z(t)$ on the finite set of states $E$. This set $E$ will be described in more details in each one  of the following subsections since it depends on the model.

 At any given time $t$, let $\mu_t(z)$ denote the proportion of investors in state $z \in E$, i.e. for each $t \geq 0$, $\mu_t$ is a probability law on $E$.

Let $m_i$ denote the proportion of asset $i$, for all $i \in \mathcal{I}$.

\subsection{Non-segmented markets}

In this simpler model, we recall that we do not track the particular asset an investor wants to buy when entering the market. Let $l$ and $h$ denote respectively a low liquidity and a high liquidity type and let $o$ and $n$ denote respectively whether an investor owns or does not own an asset. Then, the set of investors' states is fully described as follows: $E = \{(l,n),(h,n),(hi,o),(li,o)\}_{i\in\mathcal{I}}$.

As we said earlier, we do not assume  the eagerness of investors is the same when they own the asset and when they don't. For investors not-owning an asset, let us denote the switching intensity from low-type to high-type by $\gamma_u$ and conversely the  switching intensity from high-type to low-type by $\gamma_d$. For investors owning asset $i$, we will denote the switching intensity from low-type to high-type by $\gamma_{ui}$ and conversely the  switching intensity from high-type to low-type by $\gamma_{di}$. In addition, investors meet each other at rate $\lambda_i$, and an exchange of the asset occurs when an investor of type $(li,o)$ (owns asset $i$ but has a low liquidity state) meets one of type $(h,n)$ (does not own an asset but has a high interest for acquiring one).

Hence, the dynamical system describing the evolution of the proportions of investors in a given state is the following system of $2K + 2$ equations with $K+1$ constraints for $\mu_t(z)$ for each $z\in E$:
\begin{align}
	\dot{\mu}_t(h,n) &= - \mu_t(h,n)\sum_{i\in\mathcal{I}}\lambda_i\mu_t(li,o) + \gamma_u \mu_t(l,n) - \gamma_d\mu_t(h,n)\label{eq1}\\
	\dot{\mu}_t(l,n) &=  \mu_t(h,n)\sum_{i\in\mathcal{I}}\lambda_i\mu_t(li,o) - \gamma_u \mu_t(l,n) + \gamma_d\mu_t(h,n)\label{eq2}\\
	\dot{\mu}_t(hi,o) &= \lambda_i \mu_t(h,n)\mu_t(li,o) + \gamma_{ui}\mu_t(li,o) - \gamma_{di}\mu_t(hi,o), \ \forall i\in\mathcal{I}\label{eqs3}\\
	\dot{\mu}_t(li,o) &= -\lambda_i \mu_t(h,n)\mu_t(li,o) - \gamma_{ui} \mu_t(li,o) + \gamma_{di}\mu_t(hi,o), \ i\in\mathcal{I}\label{eqs4}
\end{align}
with the constraints
\begin{align*}
	\mu_t(hi,o) + \mu_t(li,o) &= m_i, \ \forall i\in\mathcal{I}\\
	\sum_{i\in\mathcal{I}} m_i  + \mu_t(h,n)  + \mu_t(l,n) &= 1
\end{align*}
This is a first generalized version of the system described in \citet{Duffie2005}.

\begin{figure}[h!]
  \centering
  \includegraphics[width=\linewidth]{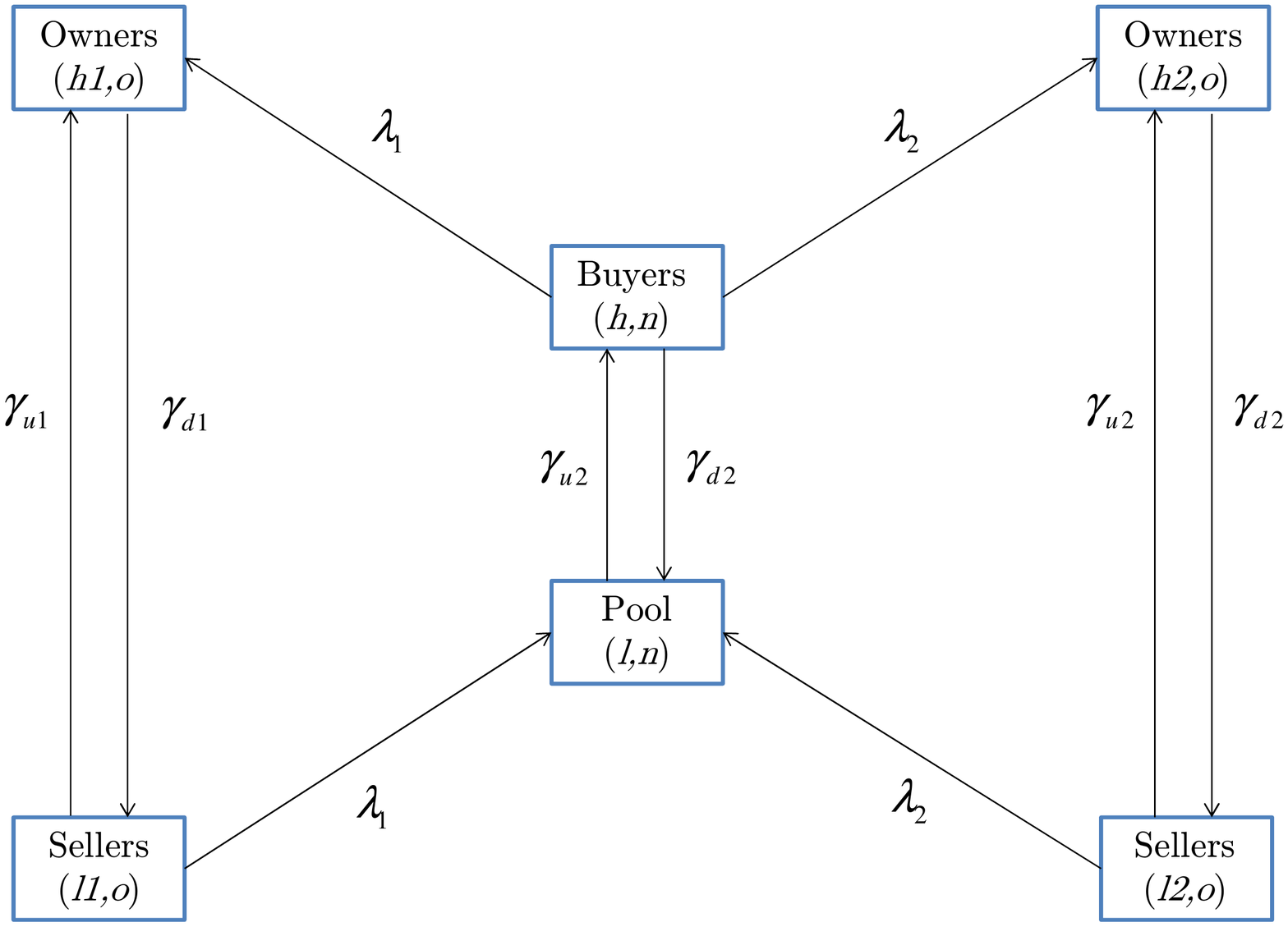}\\
  \caption{}
  \label{fig:graphNonSegm2assets}
\end{figure}

A schematic of the dynamics between investors for this class of market with two assets is illustrated on Figure \ref{fig:graphNonSegm2assets}.

Since equation (\ref{eq2}) and equation set (\ref{eqs3}) can be eliminated respectively by adding (\ref{eq2}) to (\ref{eq1}) and by adding each equation of (\ref{eqs4}) to (\ref{eqs3}), the initial system described by a set of $2+2K$ equations is reduced to the following set of $1+K$ equations:
\begin{align}\label{masterSystemEq7}
\begin{split}
	\dot{\mu}_t(h,n) &= - \mu_t(h,n)\sum_{i\in\mathcal{I}}\lambda_i\mu_t(li,o) + \gamma_u \mu_t(l,n) - \gamma_d\mu_t(h,n)\\
	\dot{\mu}_t(li,o) &= -\lambda_i \mu_t(h,n)\mu_t(li,o) - \gamma_{ui} \mu_t(li,o) + \gamma_{di}\mu_t(hi,o), \ \forall i\in\mathcal{I}
\end{split}
\end{align}
with the $1+K$ constraints
\begin{align}
	\mu_t(hi,o) + \mu_t(li,o) &= m_i, \  \forall i\in\mathcal{I}\label{constraint1}\\
	\sum_{i\in\mathcal{I}} m_i + \mu_t(h,n)  + \mu_t(l,n) &= 1\label{constraint2}
\end{align}
Note that in the first set of constraints, $m_i$ is the fraction of the investors' population holding the $i$-th asset, with $\sum_{i\in \mathcal{I}}m_i < 1$. The second constraint is the investors' proportions normalisation. Moreover, since all parameters are positive, a minus sign in the system means an exit from the state and a positive sign means an entry in the state.

The system (\ref{masterSystemEq7}) is the Master Equation. It is non-linear but there is nevertheless for each initial law $\mu_0$ a probability law  $\mathbb{P}^{\mu_0}$  on the pure jump trajectories $Z(t)$ on $E$, which has the Markov property. We do not have, however, that this law $\mathbb{P}^{\mu_0}$ is the convex combination $\sum_{z\in E}\mu_0(z)\mathbb{P}^{\delta_z}$, where $\delta_z$ are Dirac masses.
The existence of $\mathbb{P}^{\mu_0}$, on the pure jump trajectories, can be obtained by solving a martingale problem
which is built with the intensity measure, $m$, defined as follows $\forall i\in\mathcal{I}$:
\[\begin{array}{ll}
m(s,(h,n);(hi,o))=\lambda_i \mu _{s}(li,o); & m(s,(li,o);(l,n))=\lambda_i \mu
_{s}(h,n); \\
m(s,(li,o);(hi,o))=\gamma _{ui}; & m(s,(hi,o);(li,o))=\gamma _{di}; \\
m(s,(l,n);(h,n))=\gamma _{u}; & m(s,(h,n);(l,n))=\gamma _{d};
\end{array}\]
for $s \in [t,\infty)$, other terms being $0$.
This intensity measure satisfies the conditions of Theorem 2.1,
page 216, of \citet{Stroock1975}. So, once we have solved
the ODE system, for each initial condition $\mu_0$, we see that there exists a probability
measure $\mathbb{P}^{\mu_0}$. The fact that this law is supported by the set of pure jump trajectories can
be proved as in Lemma 1, page 588, of \citet{Sznitman1984}. It is such a description that we use below to obtain an
expression for the intrinsic value associated to the state of an investor at
each time. Using the properties of this expression we can then evaluate the
directly negotiated prices among investors in our relatively opaque market. One can also consult Appendix I of \citet{Duffie2001} for a review of the basic theory of intensity-based models.

It is worth noticing that the laws $\mathbb{P}^{\mu_0}$ can be obtained by a functional law of large numbers as in \citet{Ferland2008} or by rewriting the system with the help of a single kernel and then using Theorem 1 of \citet{Belanger2013}.

\citet{Weill2002} proposed a similar system with the assumption that the eagerness is the same for all assets.

\subsection{Partially segmented markets}

In this class of models, buyers who do not hold an asset enter the market with a specific asset they want to purchase. Hence, the  set of investors' type is given by $E = \{(l,n),(hi,o),(hi,n),(li,o)\}_{i\in\mathcal{I}}$. As before, the first letter designates the investor's intrinsic liquidity state and the second letter designates whether the investor owns the asset or not.

In this case, the eagerness' parametrization is the following: If an investor initially does not own any asset and is a low-type, the switching intensity of becoming a high-type is $\widetilde{\gamma}_{ui}$ and it now depends on the asset type. If he initially does not own any asset but is a high-type, he will seek to buy a specific asset $i$ and his switching intensity of becoming a low-type is $\widetilde{\gamma}_{di}$ and it now also depends on the asset type. However, if an investor initially is owning that specific asset $i$ and is a high-type (that is, he wants to keep his asset), the switching intensity of becoming a low-type is $\gamma_{di}$. If he initially owns a specific asset $i$ but is a low-type, the switching intensity of becoming a high-type is $\gamma_{ui}$. In addition, investors meet each other at rate $\lambda_i$, but an exchange of the asset occurs only if an investor of type $(li,o)$ meets one of type $(hi,n)$.

 Hence, we have the following dynamical system of investors' type proportions measure $\mu_t(z)$ for each $z\in E$, which consists of $3K+1$ equations with $K+1$ constraints:
\begin{eqnarray}
	\dot{\mu}_t(hi,n) &=& -\lambda_i \mu_t(hi,n)\mu_t(li,o) + \widetilde{\gamma}_{ui} \mu_t(l,n) - \widetilde{\gamma}_{di}\mu_t(hi,n), \ \forall i\in\mathcal{I}\label{peq1}\\
	\dot{\mu}_t(l,n) &=&  \sum_{i\in\mathcal{I}} \lambda_i \mu_t(hi,n) \mu_t(li,o)- \sum_{i\in\mathcal{I}}\widetilde{\gamma}_{ui} \mu_t(l,n) + \sum_{i\in\mathcal{I}}\widetilde{\gamma}_{di}\mu_t(hi,n)\label{peq2}\\
	\dot{\mu}_t(hi,o) &=& \lambda_i \mu_t(hi,n)\mu_t(li,o) + \gamma_{ui}\mu_t(li,o) - \gamma_{di}\mu_t(hi,o), \ \forall i\in\mathcal{I}\label{peq3}\\
	\dot{\mu}_t(li,o) &=& -\lambda_i \mu_t(hi,n)\mu_t(li,o) - \gamma_{ui} \mu_t(li,o) + \gamma_{di}\mu_t(hi,o), \ \forall i\in\mathcal{I}\label{peq4}
\end{eqnarray}
with the constraints
\begin{align*}
	\mu_t(hi,o) + \mu_t(li,o) &= m_i, \ \forall i\in\mathcal{I}\\
	\sum_{i\in\mathcal{I}}m_i  + \sum_{i\in\mathcal{I}} \mu_t(hi,n)  + \mu_t(l,n) &= 1
\end{align*}

\begin{figure}[h!]
  \centering
  \includegraphics[width=\linewidth]{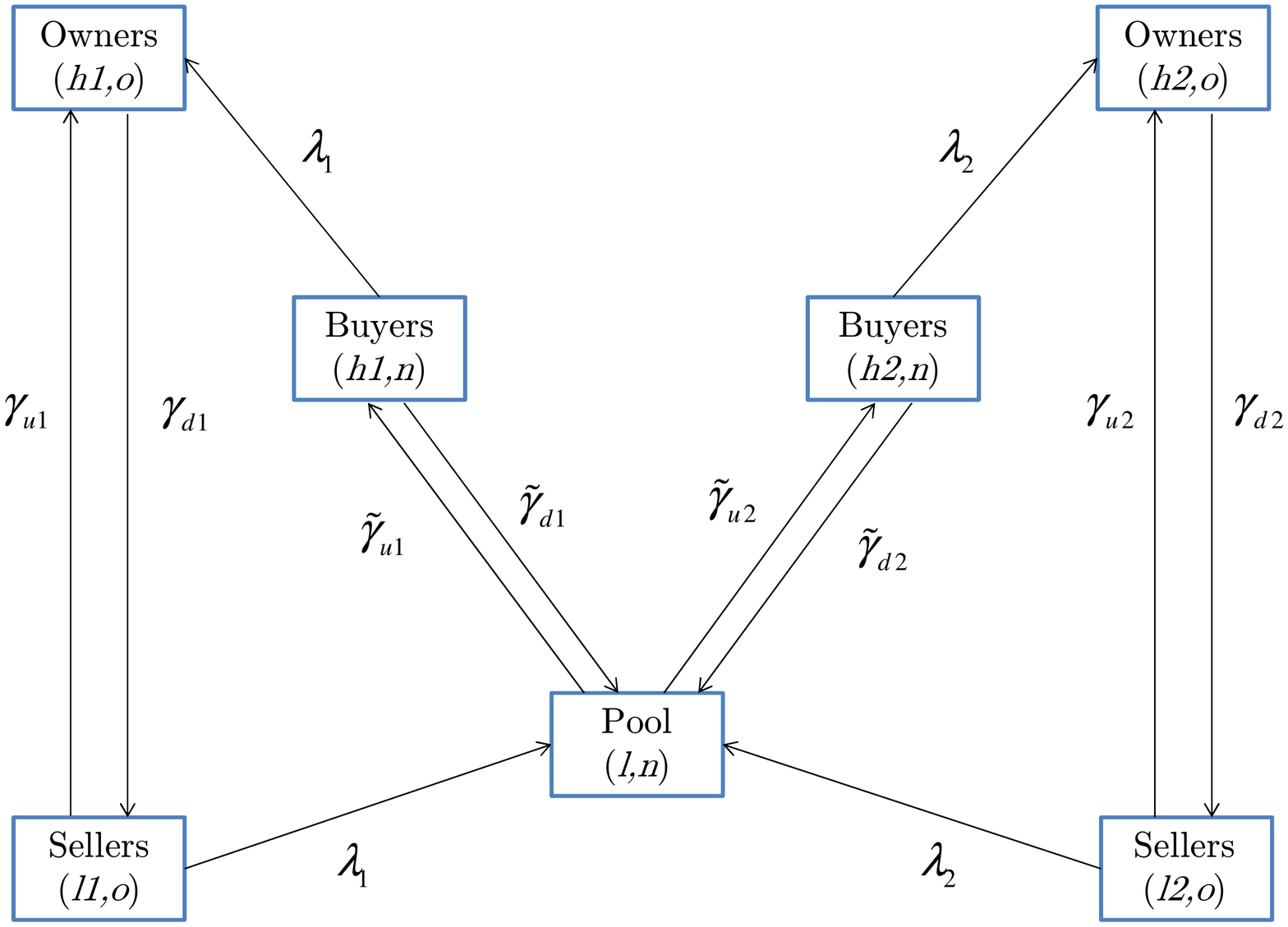}\\
  \caption{}
  \label{fig:graphSegm2assets}
\end{figure}

A schematic for the dynamics between investors for this class of models in a two assets-market ($K=2$) is illustrated on Figure \ref{fig:graphSegm2assets}.

Note that equation (\ref{peq2}) of the previous system can be eliminated by adding each equation of (\ref{peq1}) to (\ref{peq2}). Similarly, each equation of (\ref{peq3}) can be eliminated by adding it to the corresponding equation of (\ref{peq4}). The system is then reduced to the following system of $2K$ equations:
\begin{align}\label{eq:pmasterSystem}
\begin{split}
	\dot{\mu}_t(hi,n) &= -\lambda_i \mu_t(hi,n)\mu_t(li,o) + \widetilde{\gamma}_{ui} \mu_t(l,n) - \widetilde{\gamma}_{di}\mu_t(hi,n), \ \forall i\in\mathcal{I}\\
	\dot{\mu}_t(li,o) &=  -\lambda_i \mu_t(hi,n)\mu_t(li,o) - \gamma_{ui}\mu_t(li,o) + \gamma_{di}\mu_t(hi,o), \ \forall i\in\mathcal{I}
\end{split}
\end{align}
with the $1+K$ constraints
\begin{align}
	\mu_t(hi,o) + \mu_t(li,o) &= m_i, \ \forall i\in\mathcal{I}\label{pconstraint1}\\
	\sum_{i\in\mathcal{I}} m_i + \sum_{i\in\mathcal{I}} \mu_t(hi,n)  + \mu_t(l,n) &= 1\label{pconstraint2}
\end{align}

The system  (\ref{eq:pmasterSystem}) is our Master Equation and we define the intensity measure, $m$, as follows $\forall i\in\mathcal{I}$:
\[\begin{array}{ll}
m(s,(hi,n);(hi,o))=\lambda_i \mu _{s}(li,o); & m(s,(li,o);(l,n))=\lambda_i \mu
_{s}(hi,n); \\
m(s,(li,o);(hi,o))=\gamma_{ui}; & m(s,(hi,o);(li,o))=\gamma_{di}; \\
m(s,(l,n);(hi,n))=\widetilde{\gamma}_{ui}; & m(s,(hi,n);(l,n))=\widetilde{\gamma}_{di};
\end{array}\]
for $s \in [t,\infty)$, other terms being $0$.

\citet{Vayanos2007} proposed a similar two asset market.

\section{The steady state of ODE systems}

We have a steady state when the left hand side of our systems (\ref{masterSystemEq7}) and (\ref{eq:pmasterSystem}) are equal to zero. That is, when there is no longer dependence on time.

\subsection{Non-segmented markets}

Here, we need to solve the following system of equations:
\begin{align}
	0 &= -\mu(h,n)\sum_{i\in\mathcal{I}}\lambda_i\mu(li,o) + \gamma_u \mu(l,n) - \gamma_d\mu(h,n)\label{solveEq1}\\
0 &= -\lambda_i \mu(h,n)\mu(li,o) - \gamma_{ui} \mu(li,o) + \gamma_{di}\mu(hi,o), \ \forall \in\mathcal{I}\label{solveEq2}
\end{align}
First, note that we can eliminate $\mu(l,n)$ in (\ref{solveEq1}) by using the constraint equation (\ref{constraint2}). Thus, (\ref{solveEq1}) becomes
\begin{align}
	0&= - \mu(h,n)\sum_{i\in\mathcal{I}}\lambda_i\mu(li,o) + \gamma_u \left(1 - \sum_{i\in\mathcal{I}} m_i - \mu(h,n) \right)- \gamma_d\mu(h,n)\nonumber\\
		 &= - \mu(h,n)\sum_{i\in\mathcal{I}}\lambda_i\mu(li,o) + \gamma_u \left(1 - \sum_{i\in\mathcal{I}} m_i \right)  - \gamma \mu(h,n) \label{solveEq1prime}
\end{align}
where $\gamma \triangleq \gamma_d + \gamma_u$. Moreover, to simplify the last equation, we then substract the $K$ equations of (\ref{solveEq2}) to (\ref{solveEq1prime}) to have
\begin{align}
	0&= - \mu(h,n)\sum_{i\in\mathcal{I}}\lambda_i\mu(li,o) + \sum_{i\in\mathcal{I}}\left[\lambda_i\mu(h,n)\mu(li,o) + \gamma_{ui}\mu(li,o) - \gamma_{di}\mu(hi,o)\right]\nonumber\\
			&\qquad + \gamma_u \left(1 - \sum_{i=1}^K m_i \right)  - \gamma \mu(h,n)\nonumber\\
			 &=  \sum_{i\in\mathcal{I}} \gamma_{ui}\mu(li,o) - \sum_{i\in\mathcal{I}} \gamma_{di}\mu(hi,o)+ \gamma_u \left(1 - \sum_{i\in\mathcal{I}} m_i \right)  - \gamma \mu(h,n)\nonumber
\intertext{By using the constraint equation (\ref{constraint1}) to replace each $\mu(hi,o)$ in the last equation, we then have}
			 0&= \sum_{i\in\mathcal{I}} \gamma_{ui}\mu(li,o) - \sum_{i\in\mathcal{I}} \gamma_{di}(m_i - \mu(li,o))+ \gamma_u \left(1 - \sum_{i\in\mathcal{I}} m_i \right)  - \gamma \mu(h,n)\nonumber\\
		  &= \sum_{i\in\mathcal{I}} \gamma_i\mu(li,o) - \sum_{i\in\mathcal{I}} \gamma_{di}m_i + \gamma_u \left(1 - \sum_{i\in\mathcal{I}} m_i \right)  - \gamma \mu(h,n) \label{solveEq1prime2}
\end{align}
where $\gamma_i \triangleq \gamma_{di} + \gamma_{ui}$.

Furthermore, each of the $K$ equations in (\ref{solveEq2}) gives us the identity
\begin{align}
	\mu(li,o) &= \frac{\gamma_{di}m_i}{\lambda_i\mu(h,n)+\gamma_i} \label{eqMu-lio}
\end{align}
which can be substituted into (\ref{solveEq1prime2}) to have
\begin{align}
	F(\mu(h,n)) &\triangleq \sum_{i\in\mathcal{I}} \gamma_i\frac{\gamma_{di}m_i}{\lambda_i\mu(h,n)+\gamma_i} - \sum_{i\in\mathcal{I}} \gamma_{di}m_i + \gamma_u \left(1 - \sum_{i\in\mathcal{I}} m_i \right)  - \gamma \mu(h,n)\label{solveF}.
\end{align}
 Then, one need to solve $F(x) = 0$ for $x\triangleq \mu(h,n)$. Hence we get $\mu(h,n)$ from which we get  by (\ref{constraint1}) $\mu(l,n) = 1 - \mu(h,n) - \sum_{i\in\mathcal{I}} m_i$, each $\mu(li,o)$ by identity (\ref{eqMu-lio}) and finally, each $\mu(hi,o) = m_i - \mu(li,o)$, by (\ref{constraint2}).

The challenge here is to solve for $F(x) = 0$. First, note that we have
\begin{enumerate}
  \item $F(0) = \gamma_u \left(1-\sum_{i\in\mathcal{I}} m_i\right) > 0$ since $\gamma_u > 0$ and $\sum_{i\in\mathcal{I}} m_i < 1$;
  \item  $F\left(1-\sum_{i\in\mathcal{I}} m_i\right) < -\gamma_d\left(1-\sum_{i\in\mathcal{I}} m_i\right) < 0$;
  \item  $F(x)$ is a decreasing function for $x\geq 0$.
\end{enumerate}
So there is a positive root between 0 and $1-\sum_{i\in\mathcal{I}} m_i$ which can always be calculated numerically. Thus, there always exist a stationary solution $\mu(h,n)$ for any $K$.

\subsection{Partially segmented markets}

From our Master equation (\ref{eq:pmasterSystem}), we need to solve the following system of equations:
\begin{align}
	0 &=-\lambda_i \mu(hi,n)\mu(li,o) + \widetilde{\gamma}_{ui} \mu(l,n) - \widetilde{\gamma}_{di}\mu(hi,n), \ \forall i\in\mathcal{I} \label{eq:psolveEq1} \\
0 &= -\lambda_i \mu(hi,n)\mu(li,o) - \gamma_{ui}\mu(li,o) + \gamma_{di}\mu(hi,o), \ \forall i\in\mathcal{I}\label{eq:psolveEq2}
\end{align}
with the constraints
\begin{align}
	\mu(hi,o) + \mu(li,o) &= m_i, \ \forall i\in\mathcal{I} \label{eq:psolveConstr1} \\
	\sum_{i\in\mathcal{I}} m_i + \sum_{i\in\mathcal{I}} \mu(hi,n)  + \mu(l,n) &= 1 \label{eq:psolveConstr2}
\end{align}
Using each of the constraint (\ref{eq:psolveConstr1}) and substituting them in each equation of (\ref{eq:psolveEq2}) for $\mu(hi,o)$, we get
\begin{align*}
0 &= -\lambda_i \mu(hi,n)\mu(li,o) - \gamma_{ui}\mu(li,o) -\gamma_{di}\mu(li,o) + \gamma_{di}m_i, \ \forall i\in\mathcal{I}
\end{align*}
and thus
\begin{align}
\mu(li,o) &= \frac{\gamma_{di} m_i}{\lambda_i \mu(hi,n) + \gamma_i}, \ \forall i\in\mathcal{I} \label{eq:steadyMulio}
\end{align}
where $\gamma_i \triangleq \gamma_{ui}  + \gamma_{di}$.

Now, subtracting each \eqref{eq:psolveEq2} to each \eqref{eq:psolveEq1} and using constraint \eqref{eq:psolveConstr2} to substitute for $\mu(l,n)$, we get:
\begin{align*}
0 &= \widetilde{\gamma}_{ui} \left[1 - \sum_{i\in\mathcal{I}} m_i -  \sum_{i\in\mathcal{I}} \mu(hi,n) \right]   - \widetilde{\gamma}_{di}\mu(hi,n) + \gamma_{ui}\mu(li,o) - \gamma_{di}\mu(hi,o), \ \forall i\in\mathcal{I}\\
\Rightarrow \widetilde{\gamma}_i \mu(hi,n)  &=  \widetilde{\gamma}_{ui}\left(1 - \sum_{i\in\mathcal{I}} m_i \right) - \widetilde{\gamma}_{ui}  \sum_{j\neq i} \mu(hj,n ) + \gamma_{ui}\mu(li,o) - \gamma_{di}\mu(hi,o), \ \forall i\in\mathcal{I}
\end{align*}
Using constraint \eqref{eq:psolveConstr1} to substitute for $\mu(hi,o)$ and substituting \eqref{eq:steadyMulio} for $\mu(li,o)$, we finally get:
\begin{align}
\begin{split}
\mu (hi,n) &= -\frac{\widetilde{\gamma}_{ui}}{\widetilde{\gamma}_{i}} \sum_{j\neq i} \mu(hj,n)+\frac{%
\gamma_{i}\gamma_{di}m_{i}}{\widetilde{\gamma}_{i}(\lambda_i
\mu(hi,n)+\gamma_{i})}\\
&\qquad -\frac{\gamma_{di}}{%
\widetilde{\gamma} _{i}}m_{i} + \frac{\widetilde{\gamma}_{ui}}{\widetilde{\gamma}_i}\left(1-\sum_{i\in\mathcal{I}}m_{i}\right), \ \forall i\in\mathcal{I}\end{split}\label{eq:steadyMuhin}
\end{align}

\noindent Hence, we have to solve a nonlinear system of $K$ equations in $K$ unknowns $\mu(hi,n)$. Once we have solved for $\mu(hi,n)$, we can get $\mu(li,o)$ by \eqref{eq:steadyMulio} and deduce that $\mu(hi,o) = m_i - \mu(li,o)$, by \eqref{eq:psolveConstr1}, and that $\mu(l,n) = 1- \sum_{i\in\mathcal{I}}m_i - \sum_{i\in\mathcal{I}} \mu(hi,n)$, by \eqref{eq:psolveConstr2}.

Since the case $K=1$ is the same whether the market is non-segmented or partially segmented, we have the result by the previuous subsection. We will prove the case $K=2$ in the following subsection.

\subsubsection{Special case of two assets}

From (\ref{eq:steadyMuhin}) with $i\in\{1,2\}$, we get the following system to solve:
\begin{align}
\mu (h1,n) &=-\frac{\widetilde{\gamma} _{u1}}{\widetilde{\gamma} _{1}}\mu(h2,n)+\frac{%
\gamma_{1}\gamma_{d1}m_{1}}{\widetilde{\gamma}_{1}(\lambda_1
\mu(h1,n)+\gamma_{1})}+ \frac{\widetilde{\gamma}_{u1}}{\widetilde{\gamma}_1}\left(1-m_{1}-m_{2}\right)-\frac{\gamma_{d1}}{%
\widetilde{\gamma}_{1}}m_{1} \nonumber \\
\mu (h2,n) &=-\frac{\widetilde{\gamma}_{u2}}{\widetilde{\gamma}_{2}}\mu(h1,n)+\frac{%
\gamma_{2}\gamma_{d2}m_{2}}{\widetilde{\gamma}_{2}(\lambda_2
\mu(h2,n)+\gamma_{2})}+\frac{\widetilde{\gamma}_{u2}}{\widetilde{\gamma}_2}\left(1-m_{1}-m_{2}\right)-\frac{\gamma_{d2}}{%
\widetilde{\gamma}_{2}}m_{2} \nonumber
\intertext{or by rearanging terms:}
\mu (h2,n) &=-\frac{\widetilde{\gamma}_{1}}{\widetilde{\gamma}_{u1}}\mu(h1,n)+\frac{%
\gamma_{1}\gamma_{d1}m_{1}}{\widetilde{\gamma} _{u1}(\lambda_1
\mu(h1,n)+\gamma_{1})}+ 1-m_{1}-m_{2}-\frac{\gamma_{d1}}{%
\widetilde{\gamma}_{u1}}m_{1} \label{1}\\
\mu (h1,n) &=-\frac{\widetilde{\gamma} _{2}}{\widetilde{\gamma} _{u2}}\mu(h2,n)+\frac{%
\gamma_{2}\gamma_{d2}m_{2}}{\widetilde{\gamma} _{u2}(\lambda_2
\mu(h2,n)+\gamma_{2})}+1-m_{1}-m_{2}-\frac{\gamma_{d2}}{%
\widetilde{\gamma} _{u2}}m_{2}\label{2}
\end{align}
Note that the first curve \eqref{1} passes through the following two points in the set $\{(x,y)\}$, with $x\triangleq \mu(h1,n)$ and $y \triangleq \mu(h2,n)$ :
\begin{align*}
&(0,0) \leq (0 \ , \ 1-m_{1}-m_{2} )\leq (0,1)\\
&\text{and }\\
&  \left(1-m_{1}-m_{2} \ , \ -\left(1-m_{1}-m_{2}\right)\left(\frac{\widetilde{\gamma}_{1}}{\widetilde{\gamma}_{u1}}-1\right)+\left(\frac{%
\gamma_{1}}{\lambda_1 (1-m_{1}-m_{2})+\gamma_{1}}-1\right)%
\frac{\gamma_{d1}}{\widetilde{\gamma} _{u1}}m_{1}\right) \leq (1,0)
\end{align*}
because $\frac{\widetilde{\gamma}_1}{\widetilde{\gamma}_{u1}} = \frac{\widetilde{\gamma}_{u1} + \widetilde{\gamma}_{d1}}{\widetilde{\gamma}_{u1}} > 1$. By symmetry, we get that the second curve \eqref{2} passes through the points
\begin{align*}
&(0,0) \leq (1-m_{1}-m_{2} \ , \ 0)\leq (1,0)\\
&\text{and }\\
& \left(-\left(1-m_{1}-m_{2}\right)\left(\frac{\widetilde{\gamma} _{2}}{\widetilde{\gamma} _{u2}}-1\right)+\left(\frac{\gamma_{2}}{\lambda_2 (1-m_{1}-m_{2})+\gamma_{2}}-1\right)\frac{\gamma_{d2}}{\widetilde{\gamma} _{u2}}m_{2} \ , \ 1-m_{1}-m_{2}\right) \leq (0,1)
\end{align*}
because $\frac{\widetilde{\gamma}_{2}}{\widetilde{\gamma}_{u2}} > 1$.

Hence the two curves must meet in the positive unit square and we have a stationary law.

\section{Asset pricing}

  Let $C(t)$ denote the consumption process. Let $U$ a utility function and $r$, the money market interest rate (which is assumed to be constant). As before, we also have $Z(t)$, our non-homogeneous Markov chain describing investors' type (similarly to \citet{Duffie2005}). We have the following infinite-horizon expected utility maximization problem:
\begin{align}\label{eq:generalProblem}
  &\sup_{\{C(v),\theta_1(v),...,\theta_K(v)\}} \mathbb{E}\left[\int_t^{\infty}e^{-r(v-t)}U(C(v))\dd v \ | \ Z(t)=z, W(t) = w\right]
\end{align}
where the wealth process $\{W(t),  t \geq0\}$ satisfy the following equation:
\begin{align}
\begin{split}
  \dd W(t) &= rW(t)\dd t - C(t)\dd t \\
  &\qquad + \sum_{i\in\mathcal{I}}\left[\theta_i(t)\left(\delta_{hi}-\delta_{di}\mathbbm{1}_{\{Z(t)=(li,o)\}}\right)\dd t - P_i(t)\dd \theta_i(t) \right]
\end{split}
\end{align}
with $W(0) = w_0$ the initial wealth, $P_i(t)$ is the trade price between agents. $\theta_i(t)$ is the ownership process for the $i$-th asset  defined by
\begin{align}\label{eq:thetaDef}
  \theta_i(t) & =
  \left\{\begin{array}{ll}
    1, & \text{if investor owns the asset $i$, i.e. if $Z(t) \in \{(hi,o),(li,o)\}$} \\
    0, & \text{otherwise}
  \end{array}\right.
\end{align}
Note that $\dd \theta_i(t)$ here is simply a shorthand for $\theta_i(t+) - \theta_i(t-)$.

Following \citet{Duffie2005}, we will assume for simplicity that investors are risk-neutral, that is we can let $U(C(t)) = C(t)$. Hence, from (\ref{eq:generalProblem}), we define the following optimization problem:
\begin{align}\label{eq:specificProblem}
  I(t,W(t),Z(t)) &= \sup_{\{C(v),\theta_1(v),...,\theta_K(v)\}} \mathbb{E}\left[\int_t^{\infty}e^{-r(v-t)}C(v)\dd v \ | \ Z(t)=z, W(t) = w\right]
\end{align}
subject to the budget equation
\begin{align}\label{eq:specificProblemConstr}
  C(t)\dd t  & = rW(t)\dd t - \dd W(t) + \dd A(t)
\end{align}
where
\begin{align}\label{eq:dADef}
  \dd A(t) & \triangleq  \sum_{i\in\mathcal{I}}\left[\theta_i(t)\left(\delta_{hi}-\delta_{di}\mathbbm{1}_{\{Z(t)=(li,o)\}}\right)\dd t - P_i(t)\dd \theta_i(t) \right]
\end{align}

By (\ref{eq:specificProblem}) and (\ref{eq:specificProblemConstr}), we can write
\begin{align*}
   \int_t^{\infty}e^{-r(v-t)}C(v)\dd v  & = \int_t^{\infty}re^{-r(v-t)}W(v)\dd v - \int_t^{\infty}e^{-r(v-t)}\dd W(v) + \int_t^{\infty} e^{-r(v-t)}\dd A(v)\\
                        &= W(t) + \int_t^{\infty} e^{-r(v-t)}\dd A(v), \text{  by It\^{o}'s Lemma}
\end{align*}
and thus
\begin{align*}
  I(t,W(t),Z(t)) &= \sup_{\{C(v),\theta_1(v),...,\theta_K(v)\}} \mathbb{E}\left[W(t) + \int_t^{\infty} e^{-r(v-t)}\dd A(v) \ | \ Z(t)=z, W(t) = w\right]\\
            &= \sup_{\{C(v),\theta_1(v),...,\theta_K(v)\}} \left\{w + \mathbb{E}\left[ \int_t^{\infty} e^{-r(v-t)}\dd A(v) \ | \ Z(t)=z\right]\right\}
\end{align*}

\subsection{The intrinsic values $V(t,z)$}

We now want to calculate for each state $z$ the intrinsic prices at time $t$
\begin{align}\label{eq:intrinsicValue}
  V(t,z)  &\triangleq  \mathbb{E}\left[ \int_t^{\infty} e^{-r(v-t)}\dd A(v) \ | \ Z(t)=z\right].
\end{align}

Let $\tau$ be the time of the first jump in the chain $Z(t)$ after time $t$, so we can rewrite (\ref{eq:intrinsicValue}) as
\begin{align*}
  V(t,z)  &=  \mathbb{E}\left[ \int_t^{\tau} e^{-r(v-t)}\dd A(v) \ | \ Z(t)=z\right] + \mathbb{E}\left[ \int_{\tau}^{\infty} e^{-r(v-t)}\dd A(v) \ | \ Z(t)=z\right].
\end{align*}
By conditional iteration, the second term can be written as
\begin{align*}
 &\mathbb{E}\left[ \int_{\tau}^{\infty} e^{-r(v-t)}\dd A(v) \ | \ Z(t)=z\right]\\
        &\qquad= \mathbb{E}\left[ \mathbb{E}\left[  \int_{\tau}^{\infty} e^{-r(v-t)}\dd A(v) \ | \ Z(\tau) \right]\ | \ Z(t)=z\right]\\
        &\qquad= \mathbb{E}\left[ \mathbb{E}\left[  \int_{\tau}^{\infty} e^{-r((v-\tau) + (\tau- t))}\dd A(v) \ | \ Z(\tau) \right]\ | \ Z(t)=z\right]\\
        &\qquad= \mathbb{E}\left[ e^{-r(\tau-t)}\mathbb{E}\left[  \int_{\tau}^{\infty} e^{-r(v-\tau)}\dd A(v) \ | \ Z(\tau) \right]\ | \ Z(t)=z\right]\\
        &\qquad= \mathbb{E}\left[ e^{-r(\tau-t)} V(\tau,Z(\tau))  \ | \ Z(t)=z\right], \ \text{by (\ref{eq:intrinsicValue})}
\end{align*}
and thus
\begin{align*}
  V(t,z) &=  \mathbb{E}\left[ \int_t^{\tau} e^{-r(v-t)}\dd A(v) \ | \ Z(t)=z\right] + \mathbb{E}\left[ e^{-r(\tau-t)} V(\tau,Z(\tau))  \ | \ Z(t)=z\right].
\end{align*}
In the next subsections, we present the intrinsic value for each state $z\in E$ for the non-segmented and the partially segmented models. The details of calculation are in Appendix \ref{appendix:Details-V}.

\subsubsection{Intrinsic values for the  non-segmented markets}

The details of calculation for this class of models are in Appendix \ref{appendix:DetailsNonSeg-V}. The results are:
\begin{eqnarray}
V(t,(l,n))&     =& \int_{t}^{\infty}V(s,(h,n))\gamma_u \Exp{-(\gamma_u + r)(s-t)}   \dd s \label{eq:V(l,n)}\\
  V(t,(h,n))& =& \sum_{i\in\mathcal{I}} \int_{t}^{\infty} \left(V(s,(hi,o))- P_i(s)\right)\lambda_i\mu_s(li,o)\label{eq:V(h,n)}\\
   & & \times \text{exp}\left\{-\int_{t}^{s}\left(r + \gamma_{di} + \sum_{i\in\mathcal{I}}\lambda_i \mu_v(li,o)\right)\dd v\right\}\dd s\nonumber\\
   & & + \int_{t}^{\infty}V(s,(l,n))\gamma_{di} \nonumber\\
   && \times \text{exp}\left\{-\int_{t}^{s}\left(r + \gamma_{di} +  \sum_{i\in\mathcal{I}}\lambda_i \mu_v(li,o)\right)\dd v\right\} \dd s\nonumber\\
            V(t,(hi,o))&  =& \int_t^{\infty}\left(\int_t^{s} e^{-r(v-t)}\delta_{hi}\dd v \right)\gamma_{di} \text{exp}\left\{ -\gamma_{di}(s-t) \right\}\dd s \label{eq:V(hi,o)} \\
              & & +   \int_t^{\infty} V(s,(li,o)) \gamma_{di}\text{exp}\left\{ -(\gamma_{di} + r)(s-t) \right\}\dd s \nonumber\\
  V(t,(li,o)) &=& \int_t^{\infty}\left(\int_t^{s} e^{-r(v-t)}(\delta_{hi} - \delta_{di})\dd v\right)\left(\gamma_{ui}  + \lambda_i \mu_s(h,n)\right)\label{eq:V(li,o)}\\
  &&\times \text{exp}\left\{-\int_{t}^{s}\left(\gamma_{ui} + \lambda_i \mu_v(h,n)\right) \dd v\right\}\dd s\nonumber\\
 & &  +  \int_t^{\infty} V(s,(hi,o)) \gamma_{ui}\nonumber\\
 && \times\text{exp}\left\{-\int_{t}^{s}\left(\gamma_{ui} + r + \lambda_i \mu_v(h,n)\right) \dd v\right\} \dd s\nonumber\\
  &&  + \int_t^{\infty}\left( V(s,(l,n))+P_i(s)\right)\lambda_i \mu_s(h,n)\nonumber\\
  && \times \text{exp}\left\{-\int_{t}^{s}\left(\gamma_{ui} + r + \lambda_i \mu_v(h,n)\right) \dd v\right\} \dd s\nonumber
\end{eqnarray}

\subsubsection{Intrinsic values for the  partially segmented markets}

The details of calculation for this class of models are in Appendix \ref{appendix:DetailsSeg-V}. The results are:
\begin{eqnarray}
V(t,(l,n))& =&\sum_{i\in\mathcal{I}} \int_{t}^{\infty}   V(s,(hi,n))\widetilde{\gamma}_{ui} \\ \label{eq:pV(l,n)}
&& \times \text{exp}\left\{-\left(r+\sum_{i\in\mathcal{I}}\widetilde{\gamma}_{ui}\right)(s-t)\right\}\dd s\nonumber\\
  V(t,(hi,n))& =   &  \int_{t}^{\infty} \left(V(s,(hi,o))- P_i(s)\right)\lambda_i\mu_s(li,o)\label{eq:pV(hi,n)}\\
  &&\times \text{exp}\left\{-\int_{t}^{s}\left(\widetilde{\gamma}_{di} + r +\lambda_i \mu_v(li,o)\right)\dd v\right\}\dd s \nonumber\\
  &&   + \int_{t}^{\infty} V(s,(l,n))\widetilde{\gamma}_{di} \nonumber\\
  &&\times \text{exp}\left\{-\int_{t}^{s}\left(\widetilde{\gamma}_{di} + r +\lambda_i \mu_v(li,o)\right)\dd v\right\}  \dd s\nonumber\\
  V(t,(hi,o))   &=& \int_t^{\infty}\gamma_{di} \text{exp}\left\{ -\gamma_{di}(s-t) \right\}\left(\int_t^{s} e^{-r(v-t)}\delta_{hi}\dd v \right)\dd s \label{eq:pV(hi,o)} \\
              && +   \int_t^{\infty}\gamma_{di} \text{exp}\left\{ -(\gamma_{di} + r)(s-t) \right\} V(s,(li,o)) \dd s\nonumber\\
  V(t,(li,o)) &=& \int_t^{\infty}\left(\int_t^{s} e^{-r(v-t)}(\delta_{hi}- \delta_{di})\dd v\right)\left(\gamma_{ui}  + \lambda_i \mu_s(hi,n)\right)\label{eq:pV(li,o)}\\
  && \times \text{exp}\left\{-\int_{t}^{s}\left(\gamma_{ui} + \lambda_i \mu_v(hi,n)\right) \dd v\right\}\dd s\nonumber\\
  & & +  \int_t^{\infty}V(s,(hi,o))\gamma_{ui}\nonumber\\
  &&\times\text{exp}\left\{-\int_{t}^{s}\left(\gamma_{ui} + r + \lambda_i \mu_v(hi,n)\right) \dd v\right\}  \dd s\nonumber\\
  &&  + \int_t^{\infty}\left( V(s,(l,n))+P_i(s)\right)\lambda_i \mu_s(hi,n)\nonumber \\
  &&\times \text{exp}\left\{-\int_{t}^{s}\left(\gamma_{ui} + r + \lambda_i \mu_v(hi,n)\right) \dd v\right\} \dd s\nonumber
\end{eqnarray}

\subsection{ODE's for $V(t,z)$}

As we want to compute the steady prices, we first need to compute the derivative of $V(t,z)$ in time for each states $z$. We can note from the previous section that $V(t,z)$ is always of the form
\begin{align*}
  V(t,z) &= \sum_{k=1}^m\int_{t}^{\infty}g_k(z;t,s)\dd s
\end{align*}
Thus, we have
\begin{align*}
  \dot{V}(t,z) &= \frac{\partial }{\partial t} V(t,z)  \\
                &=\frac{\partial }{\partial t}\sum_{k=1}^m\int_{t}^{\infty}g_k(z;t,s)\dd s\\
                &= \sum_{k=1}^m\frac{\partial }{\partial t}\int_{t}^{\infty}g_k(z;t,s)\dd s\\
                &= \sum_{k=1}^m\left( -g_k(z;t,t) +  \int_{t}^{\infty} \frac{\partial }{\partial t} g_k(z;t,s)\dd s \right)
\end{align*}
Explicit details of these calculations are presented in Appendix \ref{appendix:Details-Vdot}.

\subsubsection{ODE's for $V(t,z)$ for the non-segmented markets}

The details of calculation for this class of models are in Appendix \ref{appendix:DetailsNonSeg-Vdot}. The results are:
\begin{eqnarray}
  \dot{V}(t,(l,n))           & =&  - V(t,(h,n))\gamma_u + (\gamma_u + r) V(t,(l,n))\label{eq:V(l,n)-point}\\
  \dot{V}(t,(h,n)) &=&  -\sum_{ i\in\mathcal{I} } \left(V(t,(hi,o))- P_i(t)\right) \lambda_i\mu_t(li,o)- \gamma_d  V(t,(l,n))\label{eq:V(h,n)-point}\\
  && + \left(\gamma_d + r + \sum_{ i\in\mathcal{I} }\lambda_i\mu_t(li,o)\right) V(t,(h,n)\nonumber\\
 \dot{V}(t,(hi,o)) &=& \left(\gamma_{di} + r\right)V(t,(hi,o)) - \gamma_{di}V(t,(li,o)) - \delta_{hi} \label{eq:V(hi,o)-point}\\
  \dot{V}(t,(li,o)) &=& \left(\gamma_{ui} + r + \lambda_i\mu_t(h,n)\right)V(t,(li,o)) - \gamma_{ui}V(t,(hi,o)) \label{eq:V(li,o)-point}\\
  && -\lambda_i \mu_t(h,n)(V(t,(l,n))+P_i(t)) - (\delta_{hi}-\delta_{di})\nonumber
\end{eqnarray}

\subsubsection{ODE's for $V(t,z)$ for the partially segmented markets}

The details of calculation for this class of models are in Appendix \ref{appendix:DetailsSeg-Vdot}. The results are:
\begin{eqnarray}
   \dot{V}(t,(l,n))           & =& -\sum_{i\in\mathcal{I}} V(t,(hi,n))\widetilde{\gamma}_{ui} + \left(r + \sum_{i\in\mathcal{I}}\widetilde{\gamma}_{ui} \right) V(t,(l,n))\label{eq:pV(l,n)-point}\\
    \dot{V}(t,(hi,n)) &=&  - \left(V(t,(hi,o))- P_i(t)\right) \lambda_i\mu_t(li,o) - V(t,(l,n)) \widetilde{\gamma}_{di}\label{eq:pV(hi,n)-point} \\
                 & & + \left(\widetilde{\gamma}_{di} + r +\lambda_i \mu_t(li,o)\right) V(t,(hi,n)\nonumber\\
  \dot{V}(t,(hi,o)) &=& \left(\gamma_{di} + r\right)V(t,(hi,o)) - \gamma_{di}V(t,(li,o)) - \delta_{hi} \label{eq:pV(hi,o)-point}\\
  \dot{V}(t,(li,o)) &=& \left(\gamma_{ui} + r + \lambda_i\mu_t(hi,n)\right)V(t,(li,o)) - \gamma_{ui}V(t,(hi,o))\label{eq:pV(li,o)-point} \\
  && -\lambda_i \mu_t(hi,n)(V(t,(l,n))+P_i(t)) - (\delta_{hi} - \delta_{di})\nonumber
\end{eqnarray}

\subsection{Equilibrium intrinsic values and prices}

The equilibrium intrinsic value are  computed by putting $\left.\frac{\partial}{\partial t}{V}(t,z)\right|_{V(z)} = 0$ for each state $z\in E$.

\subsubsection{Equilibrium prices for the non-segmented markets}

From the four equations (\ref{eq:V(l,n)-point}), (\ref{eq:V(h,n)-point}), (\ref{eq:V(hi,o)-point}) and (\ref{eq:V(li,o)-point}), we get the following system:
\begin{align*}
0  &= -\gamma_u V(h,n) + (\gamma_u + r) V(l,n)\\
0 &=  -  \sum_{i\in\mathcal{I}} \left(V(hi,o)- P_i\right)\lambda_i\mu(li,o) - \gamma_d  V(l,n) + \left(\gamma_d + r + \sum_{i\in\mathcal{I}}\lambda_i\mu(li,o)\right) V(h,n)\\
0 &= \left(\gamma_{di} + r\right)V(hi,o) - \gamma_{di}V(li,o) - \delta_{hi}, \ \forall i\in\mathcal{I}\\
  0 &= \left(\gamma_{ui} + r + \lambda_i\mu(h,n)\right)V(li,o) - \gamma_{ui}V(hi,o) \\
  &\qquad -\lambda_i \mu(h,n)(V(l,n)+P_i) - (\delta_{hi} - \delta_{di}), \ \forall i\in\mathcal{I}
\end{align*}
Rewriting this system, we get
\begin{align*}
rV(l,n)  &= \gamma_u (V(h,n) - V(l,n))\\
rV(h,n) &=  \sum_{i\in\mathcal{I}}\lambda_i\mu(li,o) \left(V(hi,o)- V(h,n) - P_i \right) + \gamma_d(V(l,n)- V(h,n))\\
rV(hi,o) &= \gamma_{di}(V(li,o) - V(hi,o)) + \delta_{hi}, \ \forall i\in\mathcal{I}\\
  rV(li,o) &= \lambda_i\mu(h,n)(V(l,n)-V(li,o)+P_i)  \\
  &\qquad + \gamma_{ui}(V(hi,o) - V(li,o)) + \delta_{hi} - \delta_{di}, \ \forall i\in\mathcal{I}
\end{align*}
Written in a similar manner as the system A5 in Appendix of \citet{Duffie2005}, but without marketmakers ($\rho = 0$), we have the following generalized system:
\begin{align*}
V(l,n)  &= \frac{\gamma_u V(h,n)}{\gamma_u + r}\\
V(h,n) &=  \frac{\sum_{i\in\mathcal{I}} \lambda_i\mu(li,o) (V(hi,o)- P_i)+\gamma_dV(l,n)}{\gamma_d + r + \sum_{i\in\mathcal{I}}\lambda_i\mu(li,o)}\\
V(hi,o) &= \frac{\gamma_{di}V(li,o) + \delta_{hi}}{\gamma_{di} +  r}, \ \forall i\in\mathcal{I}\\
  V(li,o) &=\frac{\lambda_i\mu(h,n)(V(l,n) + P_i) + \gamma_{ui}V(hi,o)+\delta_{hi}-\delta_{di}}{\gamma_{ui} + r + \lambda_i\mu(h,n)}, \ \forall i\in\mathcal{I}
\end{align*}

Now, to find the price $P_i$, we first rewrite the system in terms of the reservation prices for buyers $\Delta^h_i = V(hi,o)-V(h,n)$ and sellers $\Delta^l_i = V(li,o)-V(l,n)$. As we must have $\Delta^l_i \leq P_i \leq \Delta^h_i$, it implies that
\begin{equation}\label{eq:price}
  P_i = (1-q)\Delta^l_i + q\Delta^h_i
\end{equation}
where $q \in [0,1]$ represents the bargaining power of agents and is assumed to be the same for each asset $i\in\mathcal{I}$. Then,
\begin{align*}
rV(l,n)  &= \gamma_u (V(h,n) - V(l,n))\\
rV(h,n) &=  \sum_{i\in\mathcal{I}}\lambda_i\mu(li,o) (1-q) (\Delta^h_i - \Delta^l_i) + \gamma_d(V(l,n)- V(h,n))\\
rV(hi,o) &= \gamma_{di}(V(li,o) - V(hi,o)) + \delta_{hi}, \  \forall  i\in\mathcal{I}\\
  rV(li,o) &= \lambda_i\mu(h,n)q(\Delta^h_i - \Delta^l_i)  + \gamma_{ui}(V(hi,o) - V(li,o)) + \delta_{hi}-\delta_{di}, \  \forall  i\in\mathcal{I}
\end{align*}
Define $\Delta^0 \triangleq V(l,n)$ and $\Delta^e \triangleq V(h,n) - V(l,n)$ and rewrite the system:
\begin{align*}
r\Delta^0  &= \gamma_u \Delta^e\\
r\Delta^e &=  \sum_{i\in\mathcal{I}}\lambda_i\mu(li,o)(1- q) (\Delta^h_i - \Delta^l_i) - (\gamma_u +  \gamma_d)\Delta^e\\
r\Delta^h_i &= \gamma_{di}(\Delta^l_i - \Delta^h_i - \Delta^e) -\sum_{i\in\mathcal{I}}\lambda_i\mu(li,o) (1-q) (\Delta^h_i - \Delta^l_i) + \gamma_d\Delta^e  + \delta_{hi}, \ \forall i\in\mathcal{I}\\
  r\Delta^l_i &= \lambda_i\mu(h,n)q(\Delta^h_i - \Delta^l_i)  - \gamma_{ui}(\Delta^l_i - \Delta^h_i - \Delta^e)  - \gamma_u\Delta^e + \delta_{hi} - \delta_{di}, \ \forall i\in\mathcal{I}
\end{align*}
which is a linear system of $2K+2$ equations in $2K+2$ unknowns. If we define the vectors
\begin{equation}\label{eq:Delta}
\Delta \triangleq (\Delta_0,\Delta_1,\Delta^h_1,\Delta^h_2,...,\Delta^h_K,\Delta^l_1,\Delta^l_2,...,\Delta^l_K)^T
\end{equation}
and
\begin{equation}\label{eq:Delta}
\delta \triangleq (0,0,\delta_{h1},\delta_{h2},...,\delta_{hK},\delta_{h1}-\delta_{d1},\delta_{h2}-\delta_{d2},...,\delta_{hK} - \delta_{dK})^T,
\end{equation}
it gives us the following system to solve (which is similar to the system A7 in Appendix of \citet{Duffie2005}):
\begin{equation}\label{eq:DeltaSystem}
  M\Delta = \delta
\end{equation}
where $M$ is a $(2K+2)\times(2K+2)$ coefficient matrix defined in Appendix \ref{appendix:Matrix-NonSeg}.

If $M$ is invertible, we can solve this system by computing $\Delta = M^{-1}\delta$ and then compute asset's price  using (\ref{eq:price}).

\subsubsection{Equilibrium prices for the partially segmented markets}

Thus, from the four equations (\ref{eq:pV(l,n)-point}), (\ref{eq:pV(hi,n)-point}), (\ref{eq:pV(hi,o)-point}) and (\ref{eq:pV(li,o)-point}) of $\dot{V}(t,z)$, it gives us the following system:
\begin{align*}
0  &=-\sum_{i\in\mathcal{I}} V(t,(hi,n))\widetilde{\gamma}_{ui} + \left(r +\sum_{i\in\mathcal{I}}\widetilde{\gamma}_{ui}\right) V(t,(l,n))\\
0 &=  - \lambda_i\mu(li,o) \left(V(hi,o)- P_i\right) - \widetilde{\gamma}_{di}  V(l,n) + \left(\widetilde{\gamma}_{di} + r +\lambda_i \mu(li,o)\right) V(hi,n), \ \forall i\in\mathcal{I}\\
0 &= \left(\gamma_{di} + r\right)V(hi,o) - \gamma_{di}V(li,o) - \delta_{hi}, \ \forall i\in\mathcal{I}\\
  0 &= \left(\gamma_{ui} + r + \lambda_i\mu(hi,n)\right)V(li,o) - \gamma_{ui}V(hi,o) -\lambda_i \mu(hi,n)(V(l,n)+P_i) - (\delta_{hi}-\delta_{di}), \ \forall i\in\mathcal{I}
\end{align*}
By rewriting this system, we have
\begin{align*}
rV(l,n)  &= \sum_{i\in\mathcal{I}}\widetilde{\gamma}_{ui}(V(hi,n) - V(l,n))\\
rV(hi,n) &=  \lambda_i\mu(li,o) \left(V(hi,o)- V(hi,n) - P_i \right) + \widetilde{\gamma}_{di}(V(l,n)- V(hi,n)), \ \forall i\in\mathcal{I}\\
rV(hi,o) &= \gamma_{di}(V(li,o) - V(hi,o)) + \delta_{hi}, \ \forall i\in\mathcal{I}\\
  rV(li,o) &= \lambda_i\mu(hi,n)(V(l,n)-V(li,o)+P_i) + \gamma_{ui}(V(hi,o) - V(li,o)) + \delta_{hi}- \delta_{di}, \ \forall i\in\mathcal{I}
\end{align*}

Now, to find the price $P_i$, we first rewrite the system in terms of the reservation prices for buyers $\Delta^h_i = V(hi,o)-V(hi,n)$ and sellers $\Delta^l_i = V(li,o)-V(l,n)$. As we must have $\Delta^l_i \leq P_i \leq \Delta^h_i$, it implies that
\begin{equation}\label{eq:pprice}
  P_i = (1-q)\Delta^l_i + q\Delta^h_i
\end{equation}
where $q \in [0,1]$ represents the bargaining power of agents and is assumed to be the same for each asset $i\in\mathcal{I}$. Then, we have
\begin{align*}
rV(l,n)  &= \sum_{i\in\mathcal{I}}\widetilde{\gamma}_{ui}(V(hi,n) - V(l,n))\\
rV(hi,n) &=  \lambda_i\mu(li,o) (1-q)\left(\Delta_i^h - \Delta_i^l \right) + \widetilde{\gamma}_{di}(V(l,n)- V(hi,n)), \ \forall i\in\mathcal{I}\\
rV(hi,o) &= \gamma_{di}(V(li,o)-V(hi,o)) + \delta_{hi}, \ \forall i\in\mathcal{I}\\
  rV(li,o) &= \lambda_i\mu(hi,n)q(\Delta_i^h - \Delta_i^l) + \gamma_{ui}(V(hi,o) - V(li,o)) + \delta_{hi}-\delta_{di}, \ \forall i\in\mathcal{I}
\end{align*}
Define $\Delta^0 \triangleq V(l,n)$ and $\Delta^e_i \triangleq V(hi,n) - V(l,n)$ and rewrite the system:
\begin{align*}
r\Delta^0  &= \sum_{i\in\mathcal{I}}\widetilde{\gamma}_{ui} \Delta_i^e\\
r\Delta_i^e &= \lambda_i\mu(li,o)(1- q) (\Delta^h_i - \Delta^l_i) -  \widetilde{\gamma}_{di}\Delta_i^e - \sum_{i\in\mathcal{I}}\widetilde{\gamma}_{ui} \Delta_i^e, \ \forall i\in\mathcal{I}\\
r\Delta^h_i &= \gamma_{di}(\Delta^l_i - \Delta^h_i - \Delta_i^e) -\lambda_i\mu(li,o) (1-q) (\Delta^h_i - \Delta^l_i) + \widetilde{\gamma}_{di}\Delta_i^e + \delta_{hi}, \ \forall i\in\mathcal{I}\\
  r\Delta^l_i &= \lambda_i\mu(hi,n)q(\Delta^h_i - \Delta^l_i)  - \gamma_{ui}(\Delta^l_i - \Delta^h_i - \Delta_i^e)  - \left(\sum_{i\in\mathcal{I}}\widetilde{\gamma}_{ui}\Delta_i^e\right) + \delta_{hi} - \delta_{di}, \ \forall i\in\mathcal{I}
\end{align*}
which is a linear system of $3K+1$ equations in $3K+1$ unknowns. If we define the vectors
\begin{equation}\label{eq:pDelta}
\Delta \triangleq (\Delta^0,\Delta^e_1,\Delta_2^e,...,\Delta^e_K,\Delta^h_1,\Delta^h_2,...,\Delta^h_K,\Delta^l_1,\Delta^l_2,...,\Delta^l_K)^T
\end{equation}
and
\begin{equation}\label{eq:pDelta}
\delta \triangleq (0,0,0,...0,\delta_{h1},\delta_{h2},...,\delta_{hK},\delta_{h1}-\delta_{d1},\delta_{h2}-\delta_{d2},...,\delta_{hK}-\delta_{dK})^T,
\end{equation}
it gives us the following system to solve:
\begin{equation}\label{eq:pDeltaSystem}
  M\Delta = \delta
\end{equation}
where $M$ is a $(3K+1)\times(3K+1)$ matrix defined in Appendix \ref{appendix:Matrix-Seg}.

If $M$ is invertible, we can solve this system by computing $\Delta = M^{-1}\delta$ and then compute asset's prices using (\ref{eq:pprice}).

\section{Numerical examples for markets with two assets}

This section contains a few numerical results for our two classes of models. We present these examples primarily for an illustrative purpose.

We will use (and modify) the parameters used in \citet{Duffie2012}. We also refer to the reader to this book for the empirical justification of these parameters. That is, we assumed that $\gamma_{u1} =  \gamma_{u2} = \gamma_u = 5$ and  $\gamma_{d1} =  \gamma_{d2} = \gamma_d = 0.5$ for the non-segmented class, and $\gamma_{u1} = \widetilde{\gamma}_{u1} = \gamma_{u2}= \widetilde{\gamma}_{u2} = 5$ and  $\gamma_{d1} = \widetilde{\gamma}_{d1} = \gamma_{d2}= \widetilde{\gamma}_{d2} = 0.5$ for the segmented class. We moreover assumed that $\lambda_1 = \lambda_2 = 1250$. For comparison purpose we split in two Duffie's value of $m=0.8$ and use $m_1 = m_2 = 0.4$.

We can see in  Table \ref{tab:outputs2assets}, for the non-segmented class, under these parameters, $\mu(l1,o) = \mu(l2,o) = \mu(l,o)/2$ and $\mu(h1,o) = \mu(h2,o) = \mu(h,o)/2$, where $\mu(l,o)$ and $\mu(h,o)$ are the steady states for Duffie's one-asset market.  Note also that the prices are identical and equal to the price obtained in \citet{Duffie2012}.  The steady state proportions are different for the partially segmented market because the expected return times of the states are different. For example, it is shorter to return to $(li,o)$ because $\widetilde{\gamma}_{u1} + \widetilde{\gamma}_{u2} > \gamma_u $ so $\mu(li,o)$, equal to the reciprocal of that expected return time, is greater than in the non-segmented market. Conversely, we get a smaller $\mu(hi,o)$ because of the longer cycle in the chain that passes through $(hj,o)$, $j\neq i$. In turn, the misallocation of assets, $(li,o)$, decreases slightly the steady state price (see previous scheme on Figure \ref{fig:graphSegm2assets}).

\begin{table}[h!]
\begin{tabular}{ccccccc}
\hline
Asset $i$ & $\mu(hi,n)$ &  $\mu(h,n)$ & $\mu(li,o)$ & $\mu(hi,o)$ & $\mu(l,n)$ & $P_i$ \\
  \hline\hline
\multicolumn{7}{c}{Non-segmented}\\
\hline
Asset 1  & - & 0.1118  &  0.0014 & 0.3986  & 0.0882 & 18.5451 \\
Asset 2 & - &  0.1118  &  0.0014 &  0.3986  &  0.0882 & 18.5451\\
  \hline
     $\sum_i$   &   &          & 0.0028 &0.7972 &  & \\
  \hline\hline
  \multicolumn{7}{c}{Segmented}\\
  \hline
  Asset 1 & 0.0772 & - &  0.0020 & 0.3980 &  0.0456 &  18.3930\\
Asset 2 & 0.0772 & - &  0.0020  & 0.3980 &  0.0456 & 18.3930 \\
  \hline
    $\sum_i$  &  0.1544 &  & 0.0039 & 0.7961 &  & \\
  \hline
\end{tabular}
\caption{Models outputs. Steady state proportions and equilibrium prices.}
\label{tab:outputs2assets}
\end{table}

\begin{figure}[h!]
         \centering
                 \centering
                 \includegraphics[width=\linewidth]{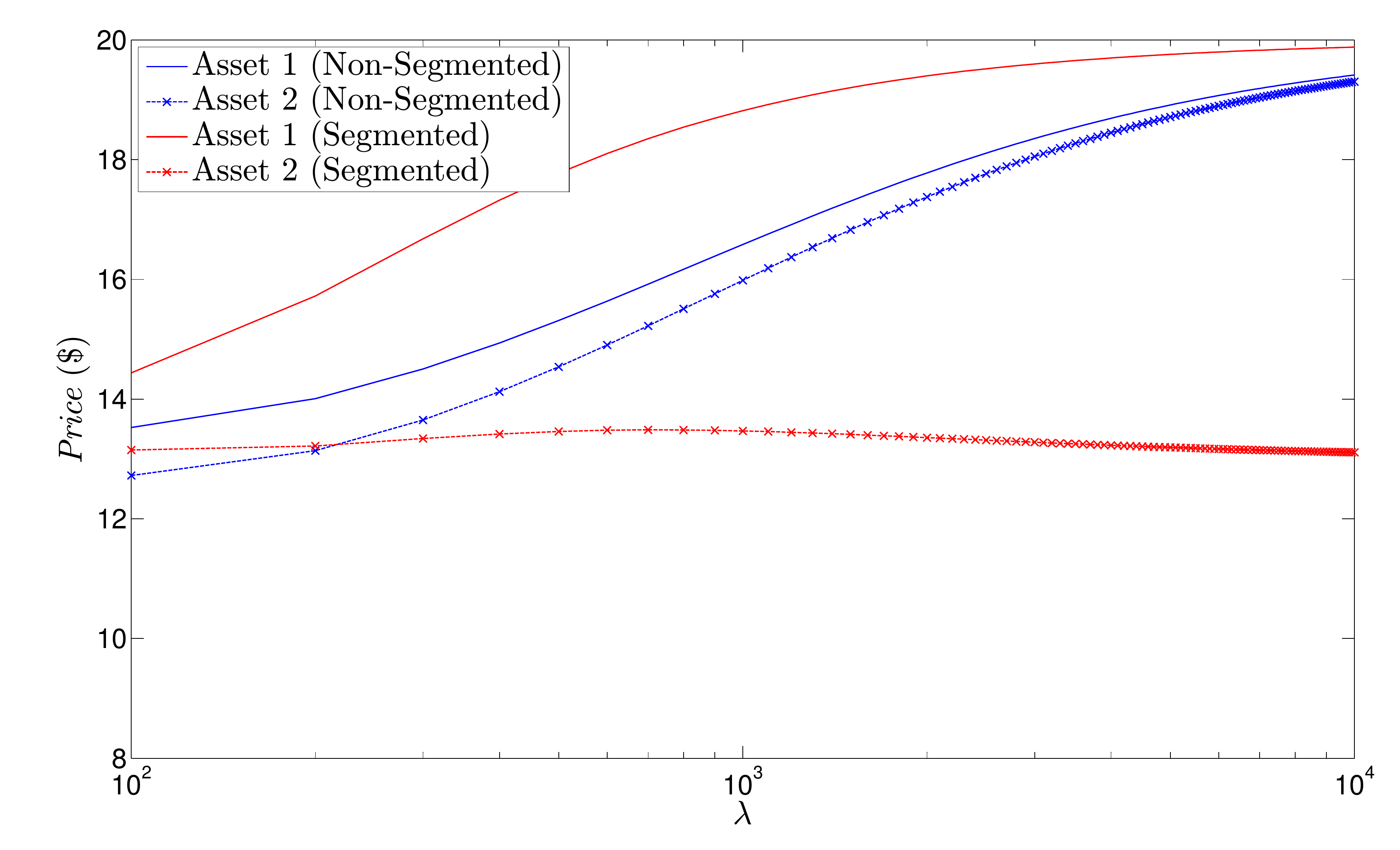}
                 \caption{$\gamma_u = \gamma_{u1} = \widetilde{\gamma}_{u1} =  \widetilde{\gamma}_{u2} = 5$, $\gamma_d = \gamma_{d2} = \widetilde{\gamma}_{d1} = \widetilde{\gamma}_{d2} =0.5$ with $\gamma_{u2} = 2\gamma_{u1}$ and $\gamma_{d2} = 2\gamma_{d1}$.}
         \label{fig:price}
\end{figure}

We now turn our attention to the sensitivity of prices with respect to $\lambda$. We still assume that $\lambda_1 = \lambda_2 = \lambda$. We can see generally that the price will tend to the perfect market price ($1/r = 20$) when frictions diminish, i.e. when $\lambda \rightarrow \infty$. The prices of the second asset in the partially segmented market exhibit a different behavior though. Its parameters $\gamma_{d2}$ and $\gamma_{u2}$ were doubled (see Figure \ref{fig:price}).

\section{Asymptotic stability}

We analyse the asymptotic stability of our ODE's systems by computing the characteristic polynomial of their Jacobian. If we can prove that all eigenvalues have negative real parts, then we have asymptotic stability (see \citet{Braun1993}).  We  do it directly, in a manner similar to that of \citet{Weill2002} for the non-segmented markets with any given number of assets. Thus, it gives us, in particular, the asymptotic stability for partially segmented markets with one asset. In order to prove the asymptotic stability for partially segmented markets with two assets, we resort to the Routh-Hurwitz criterion which gives specific conditions on the coefficients of a polynomial to ensure that the real part of all its root are  negative (see \citet{Dorf2011}).

As mentioned before, our limitation in this latter case comes from the fact
that the Routh-Hurwitz stability criterion gets very steeply harder to
verify as we increase the number of assets.

Because the Jacobian calculations involves a linear approximation of our systems close to its steady state, we prove in fact  local stability. That is, we have asymptotic stability of our systems for a subset of initial laws $\mu_0$ close to the steady state. \citet{Duffie2005} prove, more generally, the stability of their system for any initial law $\mu_0$. Their technique relies on the fact that they have a single asset and their assumption on investors' eagerness.

To simplify notations in the following subsections, we define $\widetilde{\gamma }_i%
\triangleq \widetilde{\gamma }_{ui}+\widetilde{\gamma }_{di}$, $\gamma_i \triangleq \gamma _{ui}+\gamma _{di}$ and $\gamma \triangleq \gamma_u  + \gamma_d$. Moreover, let $\xi \in \mathbb{C}$ denote the eigenvalue of the following characteristic polynomial of each system's Jacobian matrix.

\subsection{Non-segmented markets}

For this class of models, we will prove the system's stability for any $K$ assets by showing directly that all eigenvalues of the Jacobian have negative real parts.

Let $x_1 \triangleq \mu_t(l1,o)$, $x_2 \triangleq \mu_t(l2,o)$, ... , $x_K \triangleq \mu_t(lK,o)$ and $v \triangleq \mu_t(h,n)$. Then, by substituting constraint \eqref{constraint2} for $\mu_t(l,n)$ and constraints \eqref{constraint1} for each $\mu_t(hi,o)$, we can rewrite the system \eqref{masterSystemEq7} as:
\begin{eqnarray*}
x_1^{\prime } &=&-\lambda_1 x_1 v- \gamma_1 x_1 + \gamma_{d1} m_1 \\
x_2^{\prime } &=&-\lambda_2 x_2 v- \gamma_2 x_2 + \gamma_{d2} m_2 \\
              &\vdots& \\
x_K^{\prime } &=&-\lambda_K x_K v- \gamma_K x_K + \gamma_{dK} m_K \\
v^{\prime } &=&- \sum_{i\in\mathcal{I}}\lambda_i x_iv - \gamma v + \gamma_u \left(1-\sum_{i\in\mathcal{I}} m_i\right)
\end{eqnarray*}
We compute the following Jacobian matrix of the system at its steady state:
\begin{align*}
\mathbf{J} &=
\left(
  \begin{array}{ccccc}
    -\lambda_1 v  - \gamma_1 & 0 & ... & 0 & -\lambda_1 x_1  \\
    0 & -\lambda_2 v - \gamma_2 & ... & 0 & -\lambda_2 x_2\\
    \vdots & \vdots & \ddots & \vdots & \vdots \\
    0 & 0 & ... & - \lambda_K x_K- \gamma_K & -\lambda_K x_K \\
    -\lambda_1 v & -\lambda_2 v & ... & -\lambda_K v & - \left(\sum_{i\in\mathcal{I}} \lambda_i x_i\right) - \gamma \\
  \end{array}
\right)\\
&= \left(\begin{array}{ccccc}
-\gamma_1 & 0 & ... & 0 & 0\\
0 &  -\gamma_2 & ... & 0 & 0 \\
\vdots &  \vdots & \ddots & \vdots & \vdots \\
0 & 0 & ... & -\gamma_K & 0 \\
0 & 0 & ... & 0 & -\gamma
\end{array}\right) - \mathbf{D}
\end{align*}
where
\[\mathbf{D} = \left(
  \begin{array}{cc}
    D_{11}&  D_{12}\\
    D_{21} & D_{22} \\
  \end{array}
\right)\]
with $D_{11} = \text{Diag}_K(\lambda_i v)$, $D_{12} = (\lambda_1 x_1,\lambda_2 x_2,...,\lambda_K x_K)'$, $D_{21}= (\lambda_1 v, \lambda_2 v, ... , \lambda_K v)$ and $D_{22} = \sum_{i\in\mathcal{I}} \lambda_i x_i$.

Let $e \triangleq (1)_{2K+1 \times 1}$ and let $\xi \in \mathbb{C}$ be the eigenvalue for $\mathbf{J}$ associated with the eigenvector $y = (y_1,y_2)'$, where $y_1 = (y_{11},y_{22},...,y_{2K})'$ and $y_2 \in \mathbb{R}$. Then we have:
\begin{align}
 & \text{Diag}_K(\gamma_i) y_1 + D_{11}y_1 + D_{12}y_2 = -\xi y_1  \label{star1}\\
  & \gamma y_2 + D_{21}y_1 + D_{22}y_2 = -\xi y_2 \label{star2}
\end{align}
The inner product of $e'$ with \eqref{star1} gives
\begin{align}
 & \gamma_1 y_{11} + \gamma_2 y_{12} + ... + \gamma_K y_{1K} + \sum_{i\in\mathcal{I}}\lambda_i v y_{1i}+ \left(\sum_{i\in\mathcal{I}} \lambda_i x_i \right) y_2 = -\xi \sum_{i\in\mathcal{I}} y_{1i} \label{star3}
\end{align}
If we expand \eqref{star2}, we get
\begin{align}
 & \gamma y_2 + \sum_{i\in\mathcal{I}} \lambda_i v y_{1i} + \left(\sum_{i\in\mathcal{I}} \lambda_i x_i \right) y_2  = -\xi y_2  \label{star4}
\end{align}
Thus, subtracting \eqref{star4} to \eqref{star3}, we get
 \begin{align*}
 & \gamma_1 y_{11} + \gamma_2y_{12} + ... + \gamma_Ky_{1K} + \xi \sum_{i\in\mathcal{I}}y_{1i} - \gamma y_2 - \xi y_2 = 0\\
 \Rightarrow \ & e'\cdot \left[\text{Diag}_K(\gamma_i) + \xi\right]y_1 - (\gamma+\xi)y_2 = 0
\end{align*}

\noindent\textbf{Case $y_2 = 0$:} We have
\[\left(
  \begin{array}{cccc}
     -\gamma_1 & 0 & ... & 0\\
    0 & -\gamma_2 & ... & 0 \\
    \vdots & \vdots & \ddots & \vdots\\
    0 & 0 & ... & -\gamma_K
  \end{array}
\right) y_1 = \xi y_1\]
In this case, we must have $\text{Re}(\xi) < 0$ and the system is asymptotically stable.\\

\noindent\textbf{Case $y_2 \neq 0$:} We can suppose without loss of generality that $y_2 = 1$. So,
\begin{align*}
&\sum_{i\in\mathcal{I}}(\gamma_i + \xi)y_{1i} = \gamma + \xi \Rightarrow \ \sum_{i\in\mathcal{I}}  \frac{\gamma_i + \xi}{\gamma + \xi} y_{1i} = 1
\end{align*}
Then \eqref{star1} becomes
\begin{align*}
&\left(\begin{array}{c}\gamma_1y_{11}\\ \gamma_2y_{12}\\ \vdots \\ \gamma_K y_{1K}\end{array}\right) +  \left(\begin{array}{c}\lambda_1 v y_{11}\\ \lambda_2 v y_{12} \\ \vdots \\ \lambda_K v y_{1K}\end{array}\right)  +   \left(\begin{array}{c} \lambda_1 x_1\\ \lambda_2 x_2 \\ \vdots \\ \lambda_K x_K\end{array}\right)  = -\xi \left(\begin{array}{c} y_{11}\\y_{12}\\ \vdots \\ y_{1K}\end{array}\right)
 \end{align*}
 It implies that:
 \begin{eqnarray*}
y_{1i} &=& -\frac{\lambda_i x_i}{\xi + \gamma_i + \lambda_i v}, \ \forall i \in \mathcal{I}
\end{eqnarray*}
Since
\begin{align*}
  &\text{Re}\left(\sum_{i\in\mathcal{I}} \frac{\gamma_1 + \xi}{\gamma + \xi} y_{1i}\right) =1,
\end{align*}
there exists $i_0$ such that
\begin{align*}
\text{Re}(y_{1i_0}) > 0 \Leftrightarrow \text{Re}\left(\frac{1}{y_{1i_0}}\right) > 0.
\end{align*}
Then we  see that
\begin{align*}
 \text{Re}(\xi) < -(\gamma_{i_0} + \lambda_{i_0} v) < 0
\end{align*}
Thus, the system is  asymptotically stable for any number of assets.

\subsection{Partially segmented markets}

Because the partially segmented and the non-segmented markets are equivalent for one asset, the stability for one asset in this case is already proved. So we will verify  it for the case $K=2$.

 Let $x\triangleq \mu _{t}(h1,n)$, $y\triangleq \mu
_{t}(h2,n)$, $z\triangleq \mu _{t}(l1,o)$ and $v\triangleq \mu _{t}(l2,o)$. Then, as before, we
can rewrite the system as:
\begin{eqnarray*}
x^{\prime } &=&-\lambda_1 xz-\widetilde{\gamma } _{1}x-\widetilde{\gamma } _{u1}y+\widetilde{\gamma }
_{u1}(1-m_{1}-m_{2}) \\
y^{\prime } &=&-\lambda_2 yv-\widetilde{\gamma } _{2}y-\widetilde{\gamma } _{u2}x+\widetilde{\gamma }
_{u2}(1-m_{1}-m_{2}) \\
z^{\prime } &=&-\lambda_1 xz-\gamma_{1}z+\gamma_{d1}m_{1} \\
v^{\prime } &=&-\lambda_2 yv-\gamma_{2}v+\gamma_{d2}m_{2}
\end{eqnarray*}
We compute the following Jacobian matrix of the system at its steady state:
\begin{align*}
\mathbf{J} &=
\left(
  \begin{array}{cccc}
    -\lambda_1 z - \widetilde{\gamma }_1 & -\widetilde{\gamma }_{u1} & -\lambda_1 x & 0 \\
    -\widetilde{\gamma }_{u2} & -\lambda_2 v - \widetilde{\gamma }_2 & 0 & -\lambda_2 y \\
    -\lambda_1 z & 0 & -\lambda_1 x - \gamma_1 & 0 \\
    0 &  -\lambda_2 v & 0 & -\lambda_2 y - \gamma_2 \\
  \end{array}
\right)
\end{align*}

\noindent The characteristic polynomial of $\mathbf{J}$ is
\begin{align*}
  \text{det}(\mathbf{J}-\xi \mathbf{I}) &=  \text{det}\left(
  \begin{array}{cccc}
    -\lambda_1 z - \widetilde{\gamma }_1 - \xi & -\widetilde{\gamma }_{u1} & -\lambda_1 x & 0 \\
    -\widetilde{\gamma }_{u2} & -\lambda_2 v - \widetilde{\gamma }_2 - \xi & 0 & -\lambda_2 y \\
    -\lambda_1 z & 0 & -\lambda_1 x - \gamma_1  - \xi & 0 \\
    0 &  -\lambda_2 v & 0 & -\lambda_2 y - \gamma_2  - \xi\\
  \end{array}
\right)\\
&\triangleq \xi^4 + a_1 \xi^3 + a_2\xi^2 + a_3 \xi + a_4
\end{align*}
where:
\begin{eqnarray*}
a_{1} &\triangleq&  \gamma_{d1} + \gamma_{d2} + \gamma_{u1} + \gamma_{u2} + \widetilde{\gamma}_{d1} + \widetilde{\gamma}_{d2} + \widetilde{\gamma}_{u1} + \widetilde{\gamma}_{u2} + \lambda_2 (v + y) +
 \lambda_1 (x + z)\\
 a_{2} &\triangleq&  \gamma_{u1} \gamma_{u2} + \gamma_{u1} \widetilde{\gamma}_{d1} + \gamma_{u2} \widetilde{\gamma}_{d1} + \gamma_{u1} \widetilde{\gamma}_{d2} + \gamma_{u2} \widetilde{\gamma}_{d2} + \widetilde{\gamma}_{d1} \widetilde{\gamma}_{d2} + \gamma_{u1} \widetilde{\gamma}_{u1} + \gamma_{u2} \widetilde{\gamma}_{u1} \\
 && + \widetilde{\gamma}_{d2} \widetilde{\gamma}_{u1} + \gamma_{u1} \widetilde{\gamma}_{u2} + \gamma_{u2} \widetilde{\gamma}_{u2} + \widetilde{\gamma}_{d1} \widetilde{\gamma}_{u2} + \gamma_{u1} \lambda_2 v +
 \gamma_{u2} \lambda_2 v + \lambda_2 \widetilde{\gamma}_{d1} v + \lambda_2 \widetilde{\gamma}_{u1} v\\
  &&+ \gamma_{u2} \lambda_1 x + \lambda_1 \widetilde{\gamma}_{d1} x + \lambda_1 \widetilde{\gamma}_{d2} x +
 \lambda_1 \widetilde{\gamma}_{u1} x + \lambda_1 \widetilde{\gamma}_{u2} x + \lambda_1 \lambda_2 v x + \gamma_{u1} \lambda_2 y\\
  &&+ \lambda_2 \widetilde{\gamma}_{d1} y + \lambda_2 \widetilde{\gamma}_{d2} y +
 \lambda_2 \widetilde{\gamma}_{u1} y + \lambda_2 \widetilde{\gamma}_{u2} y + \lambda_1 \lambda_2 x y +
 \lambda_1 (\gamma_{u1} + \gamma_{u2} + \widetilde{\gamma}_{d2} \\
 &&+ \widetilde{\gamma}_{u2} + \lambda_2 (v + y)) z +
 \gamma_{d1} (\gamma_{d2} + \gamma_{u2} + \widetilde{\gamma}_{d1} + \widetilde{\gamma}_{d2} + \widetilde{\gamma}_{u1} + \widetilde{\gamma}_{u2} + \lambda_2 (v + y) \\
 &&+ \lambda_1 z) + \gamma_{d2} (\gamma_{u1} + \widetilde{\gamma}_{d1} + \widetilde{\gamma}_{d2} + \widetilde{\gamma}_{u1} + \widetilde{\gamma}_{u2} + \lambda_2 v + \lambda_1 (x + z))
 \end{eqnarray*}
 \begin{eqnarray*}
 a_{3} &\triangleq& \gamma_{u1} \gamma_{u2} \widetilde{\gamma}_{d1} + \gamma_{u1} \gamma_{u2} \widetilde{\gamma}_{d2} + \gamma_{u1} \widetilde{\gamma}_{d1} \widetilde{\gamma}_{d2} + \gamma_{u2} \widetilde{\gamma}_{d1} \widetilde{\gamma}_{d2} + \gamma_{u1} \gamma_{u2} \widetilde{\gamma}_{u1} + \gamma_{u1} \widetilde{\gamma}_{d2} \widetilde{\gamma}_{u1}\\
 &&+ \gamma_{u2} \widetilde{\gamma}_{d2} \widetilde{\gamma}_{u1} + \gamma_{u1} \gamma_{u2} \widetilde{\gamma}_{u2} + \gamma_{u1} \widetilde{\gamma}_{d1} \widetilde{\gamma}_{u2} + \gamma_{u2} \widetilde{\gamma}_{d1} \widetilde{\gamma}_{u2} +
  \gamma_{u1} \gamma_{u2} \lambda_2 v \\
  &&+ \gamma_{u1} \lambda_2 \widetilde{\gamma}_{d1} v + \gamma_{u2} \lambda_2 \widetilde{\gamma}_{d1} v + \gamma_{u1} \lambda_2 \widetilde{\gamma}_{u1} v +
 \gamma_{u2} \lambda_2 \widetilde{\gamma}_{u1} v + \gamma_{u2} \lambda_1 \widetilde{\gamma}_{d1} x \\
 &&+ \gamma_{u2} \lambda_1 \widetilde{\gamma}_{d2} x  + \lambda_1 \widetilde{\gamma}_{d1} \widetilde{\gamma}_{d2} x + \gamma_{u2} \lambda_1 \widetilde{\gamma}_{u1} x + \lambda_1 \widetilde{\gamma}_{d2} \widetilde{\gamma}_{u1} x + \gamma_{u2} \lambda_1 \widetilde{\gamma}_{u2} x \\
 &&+ \lambda_1 \widetilde{\gamma}_{d1} \widetilde{\gamma}_{u2} x + \gamma_{u2} \lambda_1 \lambda_2 v x + \lambda_1 \lambda_2 \widetilde{\gamma}_{d1} v x + \lambda_1 \lambda_2 \widetilde{\gamma}_{u1} v x + \gamma_{u1} \lambda_2 \widetilde{\gamma}_{d1} y \\
 && + \gamma_{u1} \lambda_2 \widetilde{\gamma}_{d2} y + \lambda_2 \widetilde{\gamma}_{d1} \widetilde{\gamma}_{d2} y + \gamma_{u1} \lambda_2 \widetilde{\gamma}_{u1} y + \lambda_2 \widetilde{\gamma}_{d2} \widetilde{\gamma}_{u1} y + \gamma_{u1} \lambda_2 \widetilde{\gamma}_{u2} y \\
 &&+ \lambda_2 \widetilde{\gamma}_{d1} \widetilde{\gamma}_{u2} y + \lambda_1 \lambda_2 \widetilde{\gamma}_{d1} x y + \lambda_1 \lambda_2 \widetilde{\gamma}_{d2} x y +
 \lambda_1 \lambda_2 \widetilde{\gamma}_{u1} x y + \lambda_1 \lambda_2 \widetilde{\gamma}_{u2} x y \\
 &&+ \lambda_1 (\gamma_{u2} (\widetilde{\gamma}_{d2} + \widetilde{\gamma}_{u2} + \lambda_2 v) + \lambda_2 (\widetilde{\gamma}_{d2} + \widetilde{\gamma}_{u2}) y + \gamma_{u1} (\gamma_{u2} + \widetilde{\gamma}_{d2} + \widetilde{\gamma}_{u2} \\
 && + \lambda_2 (v + y))) z + \gamma_{d2} (\widetilde{\gamma}_{d2} \widetilde{\gamma}_{u1} + \lambda_2 \widetilde{\gamma}_{u1} v + \lambda_1 \widetilde{\gamma}_{d2} x + \lambda_1 \widetilde{\gamma}_{u1} x + \lambda_1 \widetilde{\gamma}_{u2} x \\
 &&+ \lambda_1 \lambda_2 v x  + \widetilde{\gamma}_{d1} (\widetilde{\gamma}_{d2} + \widetilde{\gamma}_{u2} + \lambda_2 v + \lambda_1 x) +
    \lambda_1 (\widetilde{\gamma}_{d2} + \widetilde{\gamma}_{u2} + \lambda_2 v) z\\
     &&+\gamma_{u1} (\widetilde{\gamma}_{d1} + \widetilde{\gamma}_{d2} + \widetilde{\gamma}_{u1} + \widetilde{\gamma}_{u2} + \lambda_2 v + \lambda_1 z)) + \gamma_{d1} (\widetilde{\gamma}_{d1} \widetilde{\gamma}_{d2} + \widetilde{\gamma}_{d2} \widetilde{\gamma}_{u1}\\
      &&+ \widetilde{\gamma}_{d1} \widetilde{\gamma}_{u2} + \lambda_2 \widetilde{\gamma}_{d1} v + \lambda_2 \widetilde{\gamma}_{u1} v + \lambda_2 \widetilde{\gamma}_{d1} y +
    \lambda_2 \widetilde{\gamma}_{d2} y + \lambda_2 \widetilde{\gamma}_{u1} y + \lambda_2 \widetilde{\gamma}_{u2} y \\
    &&+ \lambda_1 (\widetilde{\gamma}_{d2} + \widetilde{\gamma}_{u2} + \lambda_2 (v + y)) z +
    \gamma_{d2} (\widetilde{\gamma}_{d1} + \widetilde{\gamma}_{d2} + \widetilde{\gamma}_{u1} + \widetilde{\gamma}_{u2} + \lambda_2 v + \lambda_1 z)\\
    &&+ \gamma_{u2} (\widetilde{\gamma}_{d1} + \widetilde{\gamma}_{d2} + \widetilde{\gamma}_{u1} + \widetilde{\gamma}_{u2} + \lambda_2 v + \lambda_1 z))\\
  a_{4} &\triangleq&    (\gamma_{d1} + \gamma_{u1} +
    \lambda_1 x) ((\gamma_{d2} + \gamma_{u2}) (\widetilde{\gamma}_{u1} (\widetilde{\gamma}_{d2} + \lambda_2 v) + \widetilde{\gamma}_{d1} (\widetilde{\gamma}_{d2} + \widetilde{\gamma}_{u2} + \lambda_2 v)) \\
    &&+ \lambda_2 (\widetilde{\gamma}_{d2} \widetilde{\gamma}_{u1} + \widetilde{\gamma}_{d1} (\widetilde{\gamma}_{d2} + \widetilde{\gamma}_{u2})) y) + (\gamma_{d1} +
    \gamma_{u1}) \lambda_1 ((\gamma_{d2} + \gamma_{u2}) (\widetilde{\gamma}_{d2} + \widetilde{\gamma}_{u2}\\
     &&+ \lambda_2 v) + \lambda_2 (\widetilde{\gamma}_{d2} + \widetilde{\gamma}_{u2}) y) z
\end{eqnarray*}
We can readily see that $a_{1},a_{2},a_{3},a_{4}>0.$ We need
furthermore to check that $a_{1}a_{2}a_{3}-a_{3}^{2}-a_{1}^{2}a_{4}>0$ in
order to satisfy the Routh-Hurwitz criterion which enables us to conclude
that the system is stable. This last step is done using Mathematica. We
expand the algebraic expression $a_{1}a_{2}a_{3}-a_{3}^{2}-a_{1}^{2}a_{4}$
and then simplify it to see that the result is a (very long) multiplication
and addition of positive numbers.  This shows, by Routh-Hurwitz, that the real parts of all eigenvalues are strictly
negative ensuring the asymptotic stability of the system.

 \section{Acknowledgements}

 The first author would like to thank the Universit\'{e} de Sherbrooke and its Facult\'{e} d'administration for their start-up grant.


\newpage

\appendix

\section{Calculation details}\label{appendix:Details}

\subsection{Intrinsic values $V(t,z)$}\label{appendix:Details-V}

\subsubsection{Intrinsic values for the non-segmented markets}\label{appendix:DetailsNonSeg-V} For the non-segmented markets, we have:

\noindent\textbf{Case $z=(l,n)$:} In this case, the only jump possible is towards the state $(h,n)$ with a time $\tau = \tau_h$, where $\tau_h - t \sim \mathcal{E}(\gamma_u)$. Then, for $t \leq v \leq \tau$, from (\ref{eq:thetaDef}) and (\ref{eq:dADef}), we have $\theta_i(v) = 0$, $\forall i\in\mathcal{I}$, which implies that $\dd A(v) = 0$ and
\begin{align}
  V(t,(l,n)) &= 0 +  \mathbb{E}\left[ e^{-r(\tau_h -t)} V(\tau_h,Z(\tau_h))  \ | \ Z(t)=(l,n)\right]\nonumber\\
            &= \int_{t}^{\infty}e^{-r(\tau_h-t)} V(s,(h,n))  f_{\tau_h}(s)  \dd s\nonumber\\
             &= \int_{t}^{\infty}e^{-r(s-t)} V(s,(h,n))  \gamma_u \Exp{-\gamma_u(s-t)}   \dd s\nonumber\\
              &= \int_{t}^{\infty}V(s,(h,n))\gamma_u \Exp{-(\gamma_u + r)(s-t)}  \nonumber
\end{align}

\noindent\textbf{Case $z=(h,n)$:} The next jump will be towards $(l,n)$ or $(hi,o)$, for any $i\in\mathcal{I}$. Since the investor automatically buys the first available asset, we have $\tau = \min\{\tau_l,\tau_{hi}:i\in\mathcal{I}\}$ the time until the next jump, where $\tau_l - t \sim \mathcal{E}(\gamma_d)$ but, for each $i$,  $\tau_{hi}$ has a jump intensity $\lambda_i \mu_t(li,o)$ and the following probability distribution(see Lemma 1 of \citet{Sznitman1984}):
\begin{align*}
  \mathbb{P}\{\tau_{hi} > s | Z(t) = (h,n)\} &= \Exp{-\int_{t}^{s}\lambda_i\mu_v(li,o)\dd v}, \ \text{for $s\geq t$}.
\end{align*}
In this case, for $t \leq v < \tau$, we have $\theta_{i}(v) = 0$ and $\dd A(v) = 0$, but for $v=\tau$, we have
\begin{align*}
  \theta_{i}(\tau) & =
  \left\{\begin{array}{llcl}
    1, & \text{if $\tau = \tau_{hi}$} &\Rightarrow& \dd A(\tau) =  -P_{i}(\tau_{hi})\\
    0, & \text{if $\tau = \tau_{l}$} &\Rightarrow& \dd A(\tau) = 0.
  \end{array}\right.
\end{align*}
We now compute the intrinsic value as follows:
\begin{align*}
  &V(t,(h,n))\\
  &=  \mathbb{E}\left[ \int_t^{\tau} e^{-r(s-t)}\dd A(s) \ | \ Z(t)=(hi,n)\right] + \mathbb{E}\left[ e^{-r(\tau-t)} V(\tau,Z(\tau))  \ | \ Z(t)=(hi,n)\right]\\
  &=  \sum_{i\in\mathcal{I}}\mathbb{E}\left[ e^{-r(\tau_{hi}-t)}(-P_{i}(\tau_{hi}))\mathbbm{1}_{\{\tau_l > \tau_{hi}\}\cap\{\tau_{hj}>\tau_{hi},\forall  j\neq i\}} \ | \ Z(t)=(hi,n)\right]\\
  &\qquad + 0 + \sum_{i\in\mathcal{I}}\mathbb{E}\left[ e^{-r(\tau_{hi}-t)} V(\tau_{hi},Z(\tau_{hi}))\mathbbm{1}_{\{\tau_l > \tau_{hi}\}\cap\{\tau_{hj}>\tau_{hi},\forall j\neq i\}}  \ | \ Z(t)=(hi,n)\right] \\
  & \qquad \qquad + \mathbb{E}\left[ e^{-r(\tau_{l}-t)} V(\tau_{l},Z(\tau_{l}))\mathbbm{1}_{\{\tau_l  < \tau_{hi},\forall i\}}  \ | \ Z(t)=(hi,n)\right]\\
  &= \sum_{i\in\mathcal{I}}\mathbb{E}\left[e^{-r(\tau_{hi}-t)}\left( V(\tau_{hi},(hi,o)) - P_{i}(\tau_{hi}) \right)\mathbbm{1}_{\{\tau_l > \tau_{hi}\}\cap\{\tau_{hj}>\tau_{hi},\forall  j\neq i\}} \ | \ Z(t)=(hi,n)\right]\\
  &\qquad + \mathbb{E}\left[ e^{-r(\tau_{l}-t)} V(\tau_{l},(l,n))\mathbbm{1}_{\{\tau_l  < \tau_{hi}, \forall i\}}   \ | \ Z(t)=(hi,n)\right]\\
  &=  \sum_{i\in\mathcal{I}}\int_{t}^{\infty} e^{-r(s-t)}\left(V(s,(hi,o))- P_i(s)\right) \mathbb{P}\{\tau_l > s, \tau_{hj} > s, \forall j\neq i\} f_{\tau_{hi}}(s)\dd s\\
  &\qquad + \int_{t}^{\infty}e^{-r(s-t)}V(s,(l,n))\mathbb{P}\{\tau_{hi} > s, \forall i\} f_{\tau_{l}}(s)\dd s
\end{align*}
Because we know the full intensity measure, we have that the densities are $f_{\tau_l}(s) = \gamma_d \Exp{-\gamma_d(s-t)}$ and $f_{\tau_{hi}}(s) = \lambda_i\mu_s(li,o)\Exp{-\int_{t}^{s}\lambda_i\mu_v(li,o)\dd v}$. Since the probabilities in the integrals are
\begin{align*}
  &\mathbb{P}\{\tau_l > s, \tau_{hj} > s, \forall j\neq i\} =  \Exp{-\int_{t}^{s}\left(\gamma_{di} + \sum_{\{j\in\mathcal{I}:j\neq i\}}\lambda_j \mu_v(lj,o)\right)\dd v}\\
  &\mathbb{P}\{\tau_{hi} > s, \forall i\}  =  \Exp{-\int_{t}^{s} \sum_{i\in\mathcal{I}}\lambda_i \mu_v(li,o)\dd v},
\end{align*}
then
\begin{align}
  &V(t,(h,n))\nonumber\\
       &= \sum_{i\in\mathcal{I}} \int_{t}^{\infty} e^{-r(s-t)}\left(V(s,(hi,o))- P_i(s)\right) \text{exp}\left\{-\int_{t}^{s}\left(\gamma_{di} + \sum_{\{j\in\mathcal{I}:j\neq i\}}\lambda_j \mu_v(lj,o)\right)\dd v\right\} \nonumber\\
      &\qquad\times \ \lambda_i\mu_s(li,o) \text{exp}\left\{-\int_{t}^{s} \lambda_i \mu_v(li,o)\dd v\right\}\dd s\nonumber\\
  &\qquad\qquad + \int_{t}^{\infty}e^{-r(s-t)}V(s,(l,n)) \text{exp}\left\{-\int_{t}^{s} \sum_{i\in\mathcal{I}}\lambda_i \mu_v(li,o)\dd v\right\} \gamma_{di}e^{-\gamma_{di}(s-t)}\dd s\nonumber\\
  \begin{split}
  &= \sum_{i\in\mathcal{I}} \int_{t}^{\infty} \left(V(s,(hi,o))- P_i(s)\right)\lambda_i\mu_s(li,o) \text{exp}\left\{-\int_{t}^{s}\left(r + \gamma_{di} + \sum_{i\in\mathcal{I}}\lambda_i \mu_v(li,o)\right)\dd v\right\}\dd s\\
  &\qquad + \int_{t}^{\infty}V(s,(l,n))\gamma_{di} \text{exp}\left\{-\int_{t}^{s}\left(r + \gamma_{di} +  \sum_{i\in\mathcal{I}}\lambda_i \mu_v(li,o)\right)\dd v\right\} \dd s
  \end{split}\nonumber
\end{align}

\noindent\textbf{Case $z=(hi,o)$:} The only jump possible is towards $(li,o)$ with a time $\tau = \tau_{l}$, where $\tau_l - t \sim \mathcal{E}(\gamma_{di})$. In this case, for $t \leq v \leq \tau$, we have $\theta_i(v) = 1$ and $\dd A(v) = \delta_{hi}\dd v$, so
\begin{align}
  V(t,(hi,o)) &=  \mathbb{E}\left[ \int_t^{\tau_{l}} e^{-r(v-t)}\dd A(v) \ | \ Z(t)=(hi,o)\right]\nonumber\\
   &\qquad + \mathbb{E}\left[ e^{-r(\tau_l-t)} V(\tau_l,Z(\tau_l))  \ | \ Z(t)=(hi,o)\right]\nonumber\\
                 &= \int_t^{\infty}\left(\int_t^{s} e^{-r(v-t)}\delta_{hi}\dd v \right) f_{\tau_l}(s)\dd s  +  \int_t^{\infty} e^{-r(s-t)}V(s,(li,o)) f_{\tau_l}(s)\dd s \nonumber\\
              &= \int_t^{\infty}\left(\int_t^{s} e^{-r(v-t)}\delta_{hi}\dd v \right) \gamma_{di} \text{exp}\left\{ -\gamma_{di}(s-t) \right\}\dd s\nonumber \\
              &\qquad +   \int_t^{\infty} e^{-r(s-t)}V(s,(li,o)) \gamma_{di} \text{exp}\left\{ -\gamma_{di}(s-t) \right\}\dd s \nonumber\\
  \begin{split}
              &= \int_t^{\infty}\left(\int_t^{s} e^{-r(v-t)}\delta_{hi}\dd v \right)\gamma_{di} \text{exp}\left\{ -\gamma_{di}(s-t) \right\}\dd s \\
              &\qquad +   \int_t^{\infty} V(s,(li,o)) \gamma_{di} \text{exp}\left\{ -(\gamma_{di} + r)(s-t) \right\}\dd s
  \end{split}\nonumber
\end{align}

\noindent\textbf{Case $z=(li,o)$:} The next jump will be towards $(l,n)$ or $(hi,o)$, for some $i \in \mathcal{I}$. Thus, we have $\tau = \min\{\tau_l,\tau_{hi}\}$ the time until the next jump, where $\tau_{hi} - t \sim \mathcal{E}(\gamma_{ui})$, but $\tau_{l}$ has a jump intensity $\lambda_i \mu_t(h,n)$ and the following probability distribution (see Lemma 1 of \citet{Sznitman1984}):
\begin{align*}
  \mathbb{P}\{\tau_{l} > s | Z(t) = (li,o)\} &= \text{exp}\left\{-\int_{t}^{s}\lambda_i\mu_v(h,n)\dd v\right\}, \ \text{for $s\geq t$}.
\end{align*}
In this case, for $t \leq v \leq \tau$, we have $\theta_i(v) = 1$ and $\dd A(v) = (\delta_{hi}-\delta_{di})\dd v$, but for $v=\tau$, we have
\begin{align*}
  \theta_i(\tau) & =
  \left\{\begin{array}{llcl}
    1, & \text{if $\tau = \tau_{hi}$}\\
    0, & \text{if $\tau = \tau_{l}$} &\Rightarrow& \dd A(\tau) = P_i(\tau_l).
  \end{array}\right.
\end{align*}
We now compute the intrinsic value as follows:
\begin{align*}
  V(t,(li,o)) &=  \mathbb{E}\left[ \int_t^{\tau} e^{-r(v-t)}\dd A(v) \ | \ Z(t)=(h,n)\right] \\
  &\qquad + \mathbb{E}\left[ e^{-r(\tau-t)} V(\tau,Z(\tau))  \ | \ Z(t)=(h,n)\right]\\
  &=  \mathbb{E}\left[ \int_t^{\tau} e^{-r(v-t)}(\delta_{hi} - \delta_{di})\dd v  + \left(e^{-r(\tau_l - t)}P_i(\tau_l)\right)\mathbbm{1}_{\{\tau_l < \tau_{hi}\}} \ | \ Z(t)=(li,o)\right] \\
  &\qquad + \mathbb{E}\left[ e^{-r(\tau_{hi}-t)} V(\tau_{hi},Z(\tau_{hi}))\mathbbm{1}_{\{\tau_{hi} < \tau_{l}\}}  \ | \ Z(t)=(li,o)\right] \\
  & \qquad \qquad + \mathbb{E}\left[ e^{-r(\tau_{l}-t)} V(\tau_{l},Z(\tau_{l})) \mathbbm{1}_{\{\tau_l < \tau_{hi}\}}  \ | \ Z(t)=(li,o)\right]\\
  &= \mathbb{E}\left[ \int_t^{\tau} e^{-r(v-t)} (\delta_{hi}-\delta_{di})\dd v  \ | \ Z(t)=(li,o)\right]\\
  &\qquad + \mathbb{E}\left[ e^{-r(\tau_{hi}-t)} V(\tau_{hi},(hi,o))\mathbbm{1}_{\{\tau_{hi} < \tau_{l}\}}  \ | \ Z(t)=(li,o)\right] \\
  &\qquad \qquad + \mathbb{E}\left[ e^{-r(\tau_{l}-t)}\left( V(\tau_{l},(l,n))) + P_i(\tau_l)\right)\mathbbm{1}_{\{\tau_l < \tau_{hi}\}}  \ | \ Z(t)=(li,o)\right]\\
   &= \int_t^{\infty}\left(\int_t^{s} e^{-r(v-t)}( \delta_{hi}- \delta_{di})\dd v\right)f_{\tau}(s)\dd s  +  \int_t^{\infty} V(s,(hi,o))\mathbb{P}\{\tau_{l} > s\}f_{\tau_{hi}}(s) \dd s\\
  &  \qquad + \int_t^{\infty}\left( V(s,(l,n))+P_i(s)\right)\mathbb{P}\{\tau_{hi} > s\} f_{\tau_l}(s) \dd s
\end{align*}
and thus, by knowing the density $f_\tau(s) \equiv f_{\min\{\tau_l,\tau_{hi}\}}(s)$, we have
\begin{align}
 \begin{split}
  &V(t,(li,o)) \\
  &= \int_t^{\infty}\left(\int_t^{s} e^{-r(v-t)}(\delta_{hi} - \delta_{di})\dd v\right)\left(\gamma_{ui}  + \lambda_i \mu_s(h,n)\right)\text{exp}\left\{-\int_{t}^{s}\left(\gamma_{ui} + \lambda_i \mu_v(h,n)\right) \dd v\right\}\dd s\\
  &  +  \int_t^{\infty} V(s,(hi,o)) \gamma_{ui}\text{exp}\left\{-\int_{t}^{s}\left(\gamma_{ui} + r + \lambda_i \mu_v(h,n)\right) \dd v\right\} \dd s\\
  &  + \int_t^{\infty}\left( V(s,(l,n))+P_i(s)\right)\lambda_i \mu_s(h,n)\text{exp}\left\{-\int_{t}^{s}\left(\gamma_{ui} + r + \lambda_i \mu_v(h,n)\right) \dd v\right\} \dd s
  \end{split}\nonumber
\end{align}

\subsubsection{Intrinsic values for the partially segmented markets}\label{appendix:DetailsSeg-V}For the segmented markets, we have:

\noindent\textbf{Case $z=(l,n)$:} The next jump will be towards $(hi,n)$ for any $i\in\mathcal{I}$. Thus, we have $\tau = \min\{\tau_{hi}:i\in\mathcal{I}\}$, the time until the next jump to the state $(hi,n)$, where $\tau_{hi}-t \sim \mathcal{E}(\widetilde{\gamma}_{ui})$ independently. Then, we have $\theta_i(v) = 0$ and $\dd A(v) = 0$, $\forall t \leq v \leq \tau$, and thus
\begin{align}
  V(t,(l,n)) &=  0 + \mathbb{E}\left[ e^{-r(\tau-t)} V(\tau,Z(\tau))  \ | \ Z(t)=(l,n)\right]\nonumber\\
  &= \sum_{i\in\mathcal{I}} \mathbb{E}\left[ e^{-r( \tau_{hi}-t)} V( \tau_{hi},(hi,n))\mathbbm{1}_{\{\tau_{hj} > \tau_{hi}, \forall j\neq i\}}  \ | \ Z(t)=(l,n)\right]\nonumber\\
    &= \sum_{i\in\mathcal{I}} \int_{t}^{\infty}  e^{-r(s-t)} V(s,(hi,n))\mathbb{P}\{\tau_{hj} > s, \forall j\neq i\}f_{\tau_{hi}}(s)\dd s\nonumber\\
        &= \sum_{i\in\mathcal{I}} \int_{t}^{\infty}  e^{-r(s-t)} V(s,(hi,n))\text{exp}\left\{-\sum_{\{j\in\mathcal{I}:j\neq i\}}\widetilde{\gamma}_{uj}(s-t)\right\}f_{\tau_{hi}}(s)\dd s\nonumber\\
         &= \sum_{i\in\mathcal{I}} \int_{t}^{\infty}  e^{-r(s-t)} V(s,(hi,n))\text{exp}\left\{-\sum_{\{j\in\mathcal{I}:j\neq i\}}\widetilde{\gamma}_{uj}(s-t)\right\}\widetilde{\gamma}_{ui}e^{-\widetilde{\gamma}_{ui}(s-t)}\dd s\nonumber\\
                  &= \sum_{i\in\mathcal{I}} \int_{t}^{\infty}   V(s,(hi,n))\widetilde{\gamma}_{ui}\text{exp}\left\{-\left(r+\sum_{i\in\mathcal{I}}\widetilde{\gamma}_{ui}\right)(s-t)\right\}\dd s \nonumber
\end{align}

\noindent\textbf{Case $z=(hi,n)$:} The next jump will be towards $(l,n)$ or $(hi,o)$, for some $i \in \mathcal{I}$. Thus, we have $\tau = \min\{\tau_l,\tau_{hi}\}$, the time until the next jump, where $\tau_l-t \sim \mathcal{E}(\widetilde{\gamma}_{di})$ but $\tau_{hi}$ has a jump intensity $\lambda_i \mu_t(li,o)$ with the following probability distribution (see Lemma 1 of \citet{Sznitman1984}):
\begin{align*}
  \mathbb{P}\{\tau_{hi} > s | Z(t) = (hi,n)\} &= \text{exp}\left\{-\int_{t}^{s}\lambda_i\mu_v(li,o)\dd v\right\}, \ \text{for $s\geq t$}.
\end{align*}
In this case, for $t \leq v < \tau$, we have $\theta_{i}(v) = 0$ and $\dd A(v) = 0$, but for $v=\tau$, we have
\begin{align*}
  \theta_{i}(\tau) & =
  \left\{\begin{array}{llcl}
    1, & \text{if $\tau = \tau_{hi}$} &\Rightarrow& \dd A(\tau) =  -P_{i}(\tau_{hi})\\
    0, & \text{if $\tau = \tau_{l}$} &\Rightarrow& \dd A(\tau) = 0.
  \end{array}\right.
\end{align*}
We now compute the intrinsic value as follows:
\begin{align*}
  &V(t,(hi,n))\\
  &=  \mathbb{E}\left[ \int_t^{\tau} e^{-r(v-t)}\dd A(v) \ | \ Z(t)=(hi,n)\right] + \mathbb{E}\left[ e^{-r(\tau-t)} V(\tau,Z(\tau))  \ | \ Z(t)=(hi,n)\right]\\
  &= \mathbb{E}\left[ e^{-r(\tau_{hi}-t)}(-P_{i}(\tau_{hi}))\mathbbm{1}_{\{\tau_l > \tau_{hi}\}} \ | \ Z(t)=(hi,n)\right]\\
  &\qquad +\mathbb{E}\left[ e^{-r(\tau_{hi}-t)} V(\tau_{hi},Z(\tau_{hi}))\mathbbm{1}_{\{\tau_l > \tau_{hi}\}}  \ | \ Z(t)=(hi,n)\right] \\
  & \qquad \qquad + \mathbb{E}\left[ e^{-r(\tau_{l}-t)} V(\tau_{l},Z(\tau_{l}))\mathbbm{1}_{\{\tau_l  < \tau_{hi}\}}  \ | \ Z(t)=(hi,n)\right]\\
  &= \mathbb{E}\left[e^{-r(\tau_{hi}-t)}\left( V(\tau_{hi},(hi,o)) - P_{i}(\tau_{hi}) \right)\mathbbm{1}_{\{\tau_l > \tau_{hi}\}} \ | \ Z(t)=(hi,n)\right]\\
  &\qquad + \mathbb{E}\left[ e^{-r(\tau_{l}-t)} V(\tau_{l},(l,n))\mathbbm{1}_{\{\tau_l  < \tau_{hi}\}}   \ | \ Z(t)=(hi,n)\right]\\
  &= \int_{t}^{\infty} e^{-r(s-t)}\left(V(s,(hi,o))- P_i(s)\right) \mathbb{P}\{\tau_l > s\} f_{\tau_{hi}}(s)\dd s\\
  &\qquad + \int_{t}^{\infty}e^{-r(s-t)}V(s,(l,n))\mathbb{P}\{\tau_{hi} > s\} f_{\tau_{l}}(s)\dd s\\
   &=  \int_{t}^{\infty} e^{-r(s-t)}\left(V(s,(hi,o))- P_i(s)\right)e^{-\widetilde{\gamma}_{di}(s-t)} f_{\tau_{hi}}(s)\dd s\\
  &\qquad + \int_{t}^{\infty}e^{-r(s-t)}V(s,(l,n)) \text{exp}\left\{-\int_{t}^{s} \lambda_i \mu_v(li,o)\dd v\right\}  f_{\tau_{l}}(s)\dd s\\
      &=  \int_{t}^{\infty} e^{-r(s-t)}\left(V(s,(hi,o))- P_i(s)\right) e^{-\widetilde{\gamma}_{di}(s-t)}\lambda_i \mu_s(li,o) \text{exp}\left\{-\int_{t}^{s}\lambda_i \mu_v(li,o)\dd v\right\}\dd s\\
  &\qquad + \int_{t}^{\infty}e^{-r(s-t)}V(s,(l,n)) \text{exp}\left\{-\int_{t}^{s}\lambda_i \mu_v(li,o)\dd v\right\} \widetilde{\gamma}_{di}e^{-\widetilde{\gamma}_{di}(s-t)}\dd s
 \end{align*}
 and thus
 \begin{align}
\begin{split}
   &V(t,(hi,n))  =\\
   & \ \  \int_{t}^{\infty} \left(V(s,(hi,o))- P_i(s)\right)\lambda_i\mu_s(li,o)\text{exp}\left\{-\int_{t}^{s}\left(\widetilde{\gamma}_{di} + r +\lambda_i \mu_v(li,o)\right)\dd v\right\}\dd s\\
  & \ \  + \int_{t}^{\infty} V(s,(l,n))\widetilde{\gamma}_{di} \text{exp}\left\{-\int_{t}^{s}\left(\widetilde{\gamma}_{di} + r +\lambda_i \mu_v(li,o)\right)\dd v\right\}  \dd s
\end{split}\nonumber
\end{align}

\noindent\textbf{Case $z=(hi,o)$:} In this case, the intrinsic value calculation details are identical to the case $z=(hi,o)$ in the partially segmented market.\\

\noindent\textbf{Case $z=(li,o)$:} The next jump will be towards $(l,n)$ or $(hi,o)$. Thus, we have $\tau = \min\{\tau_l,\tau_{hi}\}$ the time until the next jump, where $\tau_{hi} - t \sim \mathcal{E}(\gamma_{ui})$ but $\tau_{l}$ has a jump intensity $\lambda_i \mu_t(hi,n)$ and the following probability distribution (see Lemma 1 of \citet{Sznitman1984}):
\begin{align*}
  \mathbb{P}\{\tau_{l} > s | Z(t) = (li,o)\} &= \text{exp}\left\{-\int_{t}^{s}\lambda_i\mu_v(hi,n)\dd v\right\}, \ \text{for $s\geq t$}.
\end{align*}
In this case, for $t \leq v \leq \tau$, we have $\theta_i(v) = 1$ and $\dd A(v) = (\delta_{hi}-\delta_{di})\dd v$, but for $v=\tau$, we have
\begin{align*}
  \theta_i(\tau) & =
  \left\{\begin{array}{llcl}
    1, & \text{if $\tau = \tau_{hi}$}\\
    0, & \text{if $\tau = \tau_{l}$} &\Rightarrow& \dd A(\tau) = P_i(\tau_l).
  \end{array}\right.
\end{align*}
The remaining intrinsic value calculation details are very similar to the case $z=(li,o)$ in the partially segmented market.

\newpage
\subsection{ODE's for $V(t,z)$}\label{appendix:Details-Vdot}

\subsubsection{ODE's for $V(t,z)$ for the non-segmented markets}\label{appendix:DetailsNonSeg-Vdot}For the non-segmented markets, we have:

\noindent\textbf{Case $z=(l,n)$:} In this case, $m=1$ and we have
\begin{align*}
  g_1((l,n);t,s) &= V(s,(h,n))\gamma_u e^{-(\gamma_u + r)(s-t)}
\end{align*}
Since $ \frac{\partial }{\partial t}\ g_1((l,n);t,s) = (\gamma_u + r)g_1((l,n);t,s)$, then
\begin{align*}
  \dot{V}(t,(l,n)) &=  -g_1((l,n);t,t) +  \int_{t}^{\infty} (\gamma_u + r)g_1((l,n);t,s)\dd s \\
               &= -V(t,(h,n))\gamma_u  +(\gamma_u + r) \int_{t}^{\infty} g_1((l,n);t,s) \dd s
\end{align*}
By (\ref{eq:V(l,n)}), we then have
\begin{equation*}
  \dot{V}(t,(l,n))            = - V(t,(h,n))\gamma_u + (\gamma_u + r) V(t,(l,n))
\end{equation*}

\noindent\textbf{Case $z=(h,n)$:} In this case, $m=2$ and we have
\begin{align*}
  g_1((h,n);t,s) &=   \left(V(s,(hi,o))- P_i(s)\right) \\
  &\qquad \times \sum_{i\in\mathcal{I}}\lambda_i\mu_s(li,o)\text{exp}\left\{-\int_{t}^{s}\left(\gamma_d + r + \sum_{i\in\mathcal{I}}\lambda_i\mu_v(li,o)\right)\dd v\right\} \\
  g_2((h,n);t,s) &=  V(s,(l,n)) \gamma_d \text{exp}\left\{-\int_{t}^{s}\left(\gamma_d + r + \sum_{i\in\mathcal{I}}\lambda_i\mu_v(li,o)\right)\dd v\right\}
\end{align*}
Thus,
\begin{align*}
  \frac{\partial }{\partial t} g_1((h,n);t,s) &=  \left(\gamma_d + r + \sum_{i\in\mathcal{I}}\lambda_i\mu_t(li,o)\right)g_1((h,n);t,s)\\
  \frac{\partial }{\partial t} g_2((h,n);t,s) &=  \left(\gamma_d + r + \sum_{i\in\mathcal{I}}\lambda_i\mu_t(li,o)\right)g_2((h,n);t,s)
\end{align*}
and then
\begin{align*}
  \dot{V}(t,(h,n)) &=  -g_1((h,n);t,t) - g_2((h,n);t,t) \\
  &\qquad +   \int_{t}^{\infty}  \left(\gamma_d + r + \sum_{i\in\mathcal{I}}\lambda_i\mu_t(li,o)\right)g_1((h,n);t,s)\dd s \\
                &\qquad  +  \int_{t}^{\infty}  \left(\gamma_d + r + \sum_{i\in\mathcal{I}}\lambda_i\mu_t(li,o)\right)g_2((h,n);t,s)\dd s\\
                &=  -  \sum_{i\in\mathcal{I}}\left(V(t,(hi,o))- P_i(t)\right)\lambda_i\mu_t(li,o)  - \gamma_d  V(t,(l,n))\\
                 & \qquad + \left(\gamma_d + r +  \sum_{i\in\mathcal{I}}\lambda_i\mu_t(li,o)\right) \left(\int_{t}^{\infty}  [g_1((h,n);t,s) + g_2((h,n);t,s)]\dd s\right)
\end{align*}
By (\ref{eq:V(h,n)}),
\begin{align}
\begin{split}
  \dot{V}(t,(h,n)) &=  -\sum_{ i\in\mathcal{I} } \left(V(t,(hi,o))- P_i(t)\right) \lambda_i\mu_t(li,o)\\
  &\qquad - \gamma_d  V(t,(l,n)) + \left(\gamma_d + r + \sum_{i\in\mathcal{I} }\lambda_i\mu_t(li,o)\right) V(t,(h,n)
\end{split}\nonumber
\end{align}

\noindent\textbf{Case $z=(hi,o)$:} In this case, $m=2$ and we have
\begin{align*}
  g_1((hi,o);t,s) &= \left(\int_t^{s} e^{-r(v-t)}\delta_{hi}\dd v \right)\gamma_{di} \text{exp}\left\{ -\gamma_{di}(s-t) \right\}\\
  g_2((hi,o);t,s) &= V(s,(li,o))\gamma_{di} \text{exp}\left\{ -(\gamma_{di} + r)(s-t) \right\}
\end{align*}
Thus,
\begin{align*}
  \frac{\partial }{\partial t} g_1((hi,o);t,s) &=  \gamma_{di}\left(\gamma_{di}e^{-\gamma_{di}(s-t)}\right)\left(\int_t^s\delta_{hi}e^{-r(v-t)}\dd v\right)+ \gamma_{di}e^{-\gamma_{di}(s-t)} \frac{\partial }{\partial t}\int_t^s\delta_{hi}e^{-r(v-t)}\dd v\\
  &= \gamma_{di}g_1((hi,o);t,s) + \gamma_{di}e^{-\gamma_{di}(s-t)}\left[-\delta_{hi}e^{-r(t-t)} + \int_t^s\delta_{hi}r e^{-r(v-t)}\dd v \right]\\
  &= \gamma_{di}g_1((hi,o);t,s) - \gamma_{di}e^{-\gamma_{di}(s-t)}\delta_{hi} + r\gamma_{di}e^{-\gamma_{di}(s-t)}\int_t^s\delta_{hi} e^{-r(v-t)}\dd v\\
   &= \gamma_{di}g_1((hi,o);t,s) - \gamma_{di}e^{-\gamma_{di}(s-t)}\delta_{hi} + rg_1((hi,o);t,s)\\
   &= (\gamma_{di} +r ) g_1((hi,o);t,s) - \gamma_{di}e^{-\gamma_{di}(s-t)}\delta_{hi} \\
  \frac{\partial }{\partial t} g_2((hi,o);t,s) &=    (\gamma_{di}+r)g_2((hi,o);t,s)
\end{align*}
and then
\begin{align*}
  \dot{V}(t,(hi,o)) &=  -g_1((hi,o);t,t) - g_2((hi,o);t,t) +  \int_{t}^{\infty}\bigg[ (\gamma_{di} +r ) g_1((hi,o);t,s)\\
   &\qquad - \gamma_{di}e^{-\gamma_{di}(s-t)}\delta_{hi}\bigg]\dd s  +  \int_{t}^{\infty}  (\gamma_{di}+r)g_2((hi,o);t,s)\dd s\\
    &= 0 - \gamma_{di}V(t,(li,o)) +  (\gamma_{di}+r)\int_{t}^{\infty} [g_1((hi,o);t,s) + g_2((hi,o);t,s)]\dd s\\
     &\qquad- \int_{t}^{\infty}\gamma_{di}e^{-\gamma_{di}(s-t)}\delta_{hi}\dd s\\
         &= - \gamma_{di}V(t,(li,o)) +  (\gamma_{di}+r)\int_{t}^{\infty} [g_1((hi,o);t,s) + g_2((hi,o);t,s)]\dd s - \delta_{hi}
\end{align*}
By (\ref{eq:V(hi,o)}),
\begin{align}
  \dot{V}(t,(hi,o)) &= \left(\gamma_{di} + r\right)V(t,(hi,o)) - \gamma_{di}V(t,(li,o)) - \delta_{hi} \nonumber
\end{align}

\noindent\textbf{Case $z=(li,o)$:} In this case, $m=3$ and we have
\begin{align*}
  g_1((li,o);t,s) &= \left(\int_t^{s} e^{-r(v-t)} (\delta_{hi}-\delta_{di})\dd v\right) \left(\gamma_{ui}  + \lambda_i \mu_s(h,n)\right)\text{exp}\left\{-\int_{t}^{s}\left(\gamma_{ui} + \lambda_i \mu_v(h,n)\right) \dd v\right\}\\
  g_2((li,o);t,s) &=V(s,(hi,o))\gamma_{ui}\text{exp}\left\{-\int_{t}^{s}\left(\gamma_{ui} + r + \lambda_i \mu_v(h,n)\right) \dd v\right\} \\
  g_3((li,o);t,s) &= \left( V(s,(l,n))+P_i(s)\right)\lambda_i \mu_s(h,n)\text{exp}\left\{-\int_{t}^{s}\left(\gamma_{ui} + r + \lambda_i \mu_v(h,n)\right) \dd v\right\}
\end{align*}
Thus,
\begin{align*}
  \frac{\partial }{\partial t} g_1((li,o);t,s) &= \left(\gamma_{ui}  + \lambda_i \mu_t(h,n)\right)g_1((li,o);t,s) +  \\
  & \qquad+ \left(\gamma_{ui} \lambda_i \mu_s(h,n)\right)\text{exp}\left\{-\int_{t}^{s}\left(\gamma_{ui} + \lambda_i \mu_v(h,n)\right) \dd v\right\}\\
  &\qquad \times \left[-(\delta_{hi}-\delta_{di})e^{-r(t-t)} + \int_t^s(\delta_{hi}-\delta_{di})r e^{-r(v-t)}\dd v\right]\\
  &= \left(\gamma_{ui}  + \lambda_i \mu_t(h,n)\right)g_1((li,o);t,s) \\
  &\qquad- (\delta_{hi}-\delta_{di})\left(\gamma_{ui}  + \lambda_i \mu_s(h,n)\right)\text{exp}\left\{-\int_{t}^{s}\left(\gamma_{ui} + \lambda_i \mu_v(h,n)\right) \dd v\right\}\\
  &\qquad + r\left(\gamma_{ui}  + \lambda_i \mu_s(h,n)\right)\text{exp}\left\{-\int_{t}^{s}\left(\gamma_{ui} + \lambda_i \mu_v(h,n)\right) \dd v\right\}\int_t^s(\delta_{hi}-\delta_{di}) e^{-r(v-t)}\dd v\\
    &= \left(\gamma_{ui}  + r +  \lambda_i \mu_t(h,n)\right)g_1((li,o);t,s) \\
    &\qquad - (\delta_{hi}-\delta_{di}) \left(\gamma_{ui}  + \lambda_i \mu_s(h,n)\right)\text{exp}\left\{-\int_{t}^{s}\left(\gamma_{ui} + \lambda_i \mu_v(h,n)\right) \dd v\right\}\\
  \frac{\partial }{\partial t} g_2((li,o);t,s) &= \left(\gamma_{ui} + r + \lambda_i \mu_t(h,n)\right) g_2((li,o);t,s)\\
  \frac{\partial }{\partial t} g_3((li,o);t,s) &= \left(\gamma_{ui} + r + \lambda_i \mu_t(h,n)\right) g_3((li,o);t,s)
\end{align*}
and then
\begin{align*}
  \dot{V}(t,(li,o)) &= -g_1((li,o);t,t) - g_2((li,o);t,t) -  g_3((li,o);t,t)  \\
  &\qquad +  \int_{t}^{\infty} \bigg[\left(\gamma_{ui}  + r +  \lambda_i  \mu_t(h,n)\right)g_1((li,o);t,s) \\
  &\qquad \left. - (\delta_{hi}-\delta_{di}) \left(\gamma_{ui}  + \lambda_i \mu_s(h,n)\right)\text{exp}\left\{-\int_{t}^{s}\left(\gamma_{ui} + \lambda_i \mu_v(h,n)\right) \dd v\right\} \right]\dd s \\
  &\qquad + \int_{t}^{\infty}  \left(\gamma_{ui} + r + \lambda_i \mu_t(h,n)\right) g_2((li,o);t,s)\dd s\\
  &\qquad + \int_{t}^{\infty}  \left(\gamma_{ui} + r + \lambda_i \mu_t(h,n)\right) g_3((li,o);t,s) \dd s\\
  &= 0 - \gamma_{ui}V(t,(hi,o)) - \lambda_i \mu_t(h,n)(V(t,(l,n))+P_i(t)) \\
  &\qquad +   \left(\gamma_{ui}  + r +  \lambda_i  \mu_t(h,n)\right)\int_{t}^{\infty}[ g_1((li,o);t,s) + g_2((li,o);t,s) +g_3((li,o);t,s) ]\dd s  \\
  &\qquad - (\delta_{hi}-\delta_{di}) \int_{t}^{\infty}\left(\gamma_{ui}  + \lambda_i \mu_s(h,n)\right)\text{exp}\left\{-\int_{t}^{s}\left(\gamma_{ui} + \lambda_i \mu_v(h,n)\right) \dd v\right\}\dd s
\end{align*}
By (\ref{eq:V(li,o)}),
\begin{align}
\begin{split}
  \dot{V}(t,(li,o)) &= \left(\gamma_{ui} + r + \lambda_i \mu_t(h,n)\right)V(t,(li,o)) - \gamma_{ui}V(t,(hi,o)) \\
  &\qquad -\lambda_i \mu_t(h,n)(V(t,(l,n))+P_i(t)) - (\delta_{hi}-\delta_{di})
  \end{split}\nonumber
\end{align}

\subsubsection{ODE's for $V(t,z)$ for the partially segmented markets}\label{appendix:DetailsSeg-Vdot}For the segmented markets, we have:

\noindent\textbf{Case $z=(l,n)$:} In this case, $m=1$ and we have
\begin{align*}
  g_1((l,n);t,s) &= \sum_{i\in\mathcal{I}}V(s,(hi,n))\widetilde{\gamma}_{ui} \text{exp}\left\{-\left(r + \sum_{i\in\mathcal{I}}\widetilde{\gamma}_{ui}\right)(s-t)\right\}
\end{align*}
Since $ \frac{\partial }{\partial t}\ g_1((l,n);t,s) = \left(r + \sum_{i\in\mathcal{I}}\widetilde{\gamma}_{ui} \right)g_1((l,n);t,s)$, then
\begin{align*}
  \dot{V}(t,(l,n)) &=  -g_1((l,n);t,t) +  \int_{t}^{\infty} \left(r + \sum_{i\in\mathcal{I}}\widetilde{\gamma}_{ui} \right)g_1((l,n);t,s)\dd s \\
               &= -\sum_{i\in\mathcal{I}} V(t,(hi,n))\widetilde{\gamma}_{ui} + \left(r + \sum_{i\in\mathcal{I}}\widetilde{\gamma}_{ui}\right)\int_{t}^{\infty}g_1((l,n);t,s) \dd s
\end{align*}
By (\ref{eq:pV(l,n)}), we then have
\begin{equation}
  \dot{V}(t,(l,n))            = -\sum_{i\in\mathcal{I}} V(t,(hi,n))\widetilde{\gamma}_{ui} + \left(r+\sum_{i\in\mathcal{I}}\widetilde{\gamma}_{ui}\right) V(t,(l,n))\nonumber
\end{equation}

\noindent\textbf{Case $z=(hi,n)$:}  In this case, $m=2$ and we have
\begin{align*}
  g_1((hi,n);t,s) &=  \left(V(s,(hi,o))- P_i(s)\right)\lambda_i\mu_s(li,o)\text{exp}\left\{-\int_{t}^{s}\left(\widetilde{\gamma}_{di} + r +\lambda_i \mu_v(li,o)\right)\dd v\right\} \\
  g_2((hi,n);t,s) &=V(s,(l,n)) \widetilde{\gamma}_{di} \text{exp}\left\{-\int_{t}^{s}\left(\widetilde{\gamma}_{di} + r +\lambda_i\mu_v(li,o)\right)\dd v\right\}
\end{align*}
Thus,
\begin{align*}
  \frac{\partial }{\partial t} g_1((hi,n);t,s) &=  \left(\widetilde{\gamma}_{di} + r +\lambda_i\mu_t(li,o)\right)g_1((hi,n);t,s)\\
  \frac{\partial }{\partial t} g_2((hi,n);t,s) &=  \left(\widetilde{\gamma}_{di} + r +\lambda_i\mu_t(li,o)\right)g_2((hi,n);t,s)
\end{align*}
and then
\begin{align*}
  \dot{V}(t,(hi,n)) &=  -g_1((hi,n);t,t) - g_2((hi,n);t,t) \\
  &\qquad +   \int_{t}^{\infty}  \left(\widetilde{\gamma}_{di} + r +\lambda_i\mu_t(li,o)\right)g_1((hi,n);t,s)\dd s \\
                &\qquad  +  \int_{t}^{\infty}  \left(\widetilde{\gamma}_{di} + r +\lambda_i\mu_t(li,o)\right)g_2((hi,n);t,s)\dd s\\
                &=  -  \left(V(t,(hi,o))- P_i(t)\right)\lambda_i\mu_t(li,o)  -   V(t,(l,n))\widetilde{\gamma}_{di}\\
                 & \qquad + \left(\widetilde{\gamma}_{di} + r +\lambda_i\mu_t(li,o)\right) \left(\int_{t}^{\infty}  [g_1((hi,n);t,s) + g_2((hi,n);t,s)]\dd s\right)
\end{align*}
By (\ref{eq:pV(hi,n)}),
\begin{align}
\begin{split}
  \dot{V}(t,(hi,n)) &=  - \left(V(t,(hi,o))- P_i(t)\right) \lambda_i\mu_t(li,o) - V(t,(l,n)) \widetilde{\gamma}_{di} \\
                 & \qquad + \left(\widetilde{\gamma}_{di} + r +\lambda_i \mu_t(li,o)\right) V(t,(hi,n)
\end{split}\nonumber
\end{align}

\noindent\textbf{Case $z=(hi,o)$:} In this case, the calculation details are identical to the case $z=(hi,o)$ in the partially segmented market.\\

\noindent\textbf{Case $z=(li,o)$:} In this case, $m=3$ and we have
\begin{align*}
  g_1((li,o);t,s) &= \left(\int_t^{s} e^{-r(v-t)} (\delta_{hi}-\delta_{di})\dd v\right)\left(\gamma_{ui}  + \lambda_i \mu_s(hi,n)\right)\text{exp}\left\{-\int_{t}^{s}\left(\gamma_{ui} + \lambda_i \mu_v(hi,n)\right) \dd v\right\}\\
  g_2((li,o);t,s) &=V(s,(hi,o))\gamma_{ui}\text{exp}\left\{-\int_{t}^{s}\left(\gamma_{ui} + r + \lambda_i \mu_v(hi,n)\right) \dd v\right\} \\
  g_3((li,o);t,s) &=\left( V(s,(l,n))+P_i(s)\right)\lambda_i\mu_s(hi,n)\text{exp}\left\{-\int_{t}^{s}\left(\gamma_{ui} + r + \lambda_i \mu_v(hi,n)\right) \dd v\right\}
\end{align*}
The remaining calculation details are very similar to the case $z=(li,o)$ in the partially segmented market.

\newpage

\section{Details of the $M$ matrices}\label{appendix:Matrix}

\subsection{The non-segmented markets}\label{appendix:Matrix-NonSeg}

The $(2K+1)\times (2K+1)$ coefficients matrix in this case is defined as follows:
\begin{align*}
\hspace*{-0.5in}
&M=\left(
  \begin{array}{cccccccccc}
    r & -\gamma_u & 0 & 0 & ... & 0 & 0 & 0 & ... & 0 \\
    0 & r + \gamma & -M_1 & -M_2 & ... & -M_K & M_1 & M_2 & ... & M_K \\
    0 & -\gamma_d + \gamma_{d1} & r + \Psi_{d1} & M_2 & ... & M_K & -\Psi_{d1} & -M_2 & ... & -M_K \\
    0 & -\gamma_d + \gamma_{d2} & M_1 & r + \Psi_{d2} & ... & M_K & -M_1 & -\Psi_{d2} & ... & -M_K \\
    \vdots & \vdots & \vdots & \vdots & \ddots & \vdots & \vdots & \vdots &  \ddots &\vdots \\
    0 & -\gamma_d + \gamma_{dK} & M_1 & M_2 & ... & r + \Psi_{dK} & -M_1 & -M_2 & ... & -\Psi_{dK} \\
    0 & \gamma_u -\gamma_{u1} & -\Psi_{u1} & 0 & ... & 0 & r + \Psi_{u1} & 0 & ... & 0 \\
    0 & \gamma_u - \gamma_{u2} & 0 & -\Psi_{u2} & ... & 0 & 0 & r+\Psi_{u2} & ... & 0 \\
    \vdots & \vdots & \vdots & \vdots & \ddots & \vdots & \vdots & \vdots &  \ddots &\vdots \\
    0 & \gamma_u - \gamma_{uK} & 0 & 0 & ... & -\Psi_{uK} & 0 & 0 & ... & r + \Psi_{uK} \\
  \end{array}
\right)
\end{align*}
where $\gamma \triangleq \gamma_u + \gamma_d$, $\Psi_{ui} \triangleq \gamma_{ui} + \lambda_i\mu(h,n)q$, $\Psi_{di} \triangleq \gamma_{di} + M_i$ and $M_i \triangleq \lambda_i\mu(li,o)(1-q)$.

\subsection{The partially segmented markets}\label{appendix:Matrix-Seg}
(On next page)

\begin{landscape}
\noindent The $(3K+1)\times (3K+1)$ coefficients matrix in this case is defined as follows:
\begin{align*}
&M=\left(
  \begin{array}{ccccccccccccc}
    r & -\widetilde{\gamma}_{u1} & -\widetilde{\gamma}_{u2} & ... & -\widetilde{\gamma}_{uK} & 0 & 0 & ... & 0 & 0 & 0 & ... & 0 \\
    0 & r + \widetilde{\gamma}_1 & \widetilde{\gamma}_{u2} & ... & \widetilde{\gamma}_{uK} & -M_1 & 0 & ... & 0 & M_1 & 0 & ... & 0 \\
    0 & \widetilde{\gamma}_{u1} & r + \widetilde{\gamma}_2 & ... & \widetilde{\gamma}_{uK} & 0 & -M_2 & ... & 0 & 0 & M_2 & ... & 0 \\
    \vdots & \vdots & \vdots & \ddots & \vdots & \vdots & \vdots & \ddots & \vdots & \vdots & \vdots & \ddots & \vdots \\
    0 &\widetilde{\gamma}_{u1} & \widetilde{\gamma}_{u2} & ... &  r + \widetilde{\gamma}_K & 0 & 0 & ... & -M_K & 0 & 0 & ... & M_K \\
    0 & -\widetilde{\gamma}_{d1} + \gamma_{d1} & 0 & ... & 0 & r + \Psi_{d1} & 0 & ... & 0 & -\Psi_{d1} & 0 & ... & 0 \\
    0 & 0 & -\widetilde{\gamma}_{d2} + \gamma_{d2} & ... & 0 & 0 & r + \Psi_{d2} & ... & 0 & 0 & -\Psi_{d2} & ... & 0 \\
    \vdots & \vdots & \vdots & \ddots & \vdots & \vdots & \vdots & \ddots & \vdots & \vdots & \vdots & \ddots & \vdots \\
    0 & 0 & 0 & ... & -\widetilde{\gamma}_{dK} + \gamma_{dK} & 0 & 0 & ... & r + \Psi_{dK} & 0 & 0 & ... & -\Psi_{dK} \\
    0 &  \widetilde{\gamma}_{u1}-\gamma_{u1} & \widetilde{\gamma}_{u2} & ... & \widetilde{\gamma}_{uK} & -\Psi_{u1} & 0 & ... & 0 & r + \Psi_{u1} & 0 & ... & 0 \\
    0 & \widetilde{\gamma}_{u1} &  \widetilde{\gamma}_{u2}-\gamma_{u2} & ... &  \widetilde{\gamma}_{uK}  & 0 & -\Psi_{u2} & ... & 0 & 0 &  r + \Psi_{u2} & ... & 0 \\
    \vdots & \vdots & \vdots & \ddots & \vdots & \vdots & \vdots & \ddots & \vdots & \vdots & \vdots & \ddots & \vdots \\
    0 &  \widetilde{\gamma}_{u1}  &  \widetilde{\gamma}_{u2}  & ... &   \widetilde{\gamma}_{uK}-\gamma_{uK} & 0 & 0 & ... & -\Psi_{uK} & 0 & 0 & ... & r + \Psi_{uK} \\
  \end{array}
\right)
\end{align*}
where $\widetilde{\gamma}_i \triangleq \widetilde{\gamma}_{ui} + \widetilde{\gamma}_{di}$, $\Psi_{ui} \triangleq \gamma_{ui} + \lambda_i\mu(hi,n)q$, $\Psi_{di} \triangleq \gamma_{di} + M_i$ and $M_i \triangleq \lambda_i\mu(li,o)(1-q)$.

\end{landscape}



\bibliographystyle{imsart-nameyear}
\bibliography{bibliographyOTC}

\end{document}